\documentclass[aip,amsmath,amssymb,jcp,preprint]{revtex4-1}

\usepackage{color}
\usepackage{dcolumn}
\usepackage{graphicx}
\usepackage{tabularx}
\usepackage{enumerate}
\usepackage{dsfont}
\usepackage{mathrsfs}

\newcommand{\chng}[1]{\protect{\color{black}#1}}

\newcommand{\adcf}{ADC(2,2)$_f$}
\newcommand{\adcm}{ADC(2,2)$_m$}
\newcommand{\adcx}{ADC(2,2)$_x$}
\newcommand{\Ne}{\text{Ne}}
\newcommand{\He}{\text{He}}
\newcommand{\QP}{\mathcal{Q/P}}
\newcommand{\Psub}{\mathcal{P}}
\newcommand{\Qsub}{\mathcal{Q}}
\newcommand{\lsq}{\mathcal{L}^2}

\newcommand{\isr}[1]{|\tilde{\Psi}_{#1}\rangle}
\newcommand{\CE}[1]{|\Psi^0_{#1}\rangle}

\newcommand{\mc}[1]{\multicolumn{1}{c}{#1}}

\newcolumntype{L}[1]{>{\raggedright\arraybackslash}p{#1}}
\newcolumntype{C}[1]{>{\centering\arraybackslash}p{#1}}
\newcolumntype{R}[1]{>{\raggedleft\arraybackslash}p{#1}}

\begin{document}
\title{Fano-ADC(2,2) method for electronic decay rates} 

\author{P.\ Koloren\v{c}}
\email[]{kolorenc@mbox.troja.mff.cuni.cz}
\affiliation{Charles University, Faculty of Mathematics and Physics, Institute of Theoretical Physics, V Hole\v{s}ovi\v{c}k\'ach 2, 180 00 Prague, Czech Republic}
\author{V.\ Averbukh}
\affiliation{Blackett Laboratory, Imperial College London, London SW9 2AZ, UK}

\date{\today}

\begin{abstract}
Fano-ADC is a family of \textit{ab initio} methods for prediction of electronic decay widths in excited, singly- and doubly-ionized systems. It has been particularly successful in elucidating the geometry dependence of the inter-atomic decay widths in clusters and facilitated prediction of new electronic decay phenomena. However, the available Fano-ADC schemes are limited to the second order treatment of the initial state and first-order treatment of the final states of the decay. This confines the applicability of the Fano-ADC approach to first-order decay processes, e.g. normal but not double Auger decay, and compromises the numerical accuracy of the schemes through the unbalanced treatment of electronic correlation. Here we introduce the ADC(2,2) approximation for singly ionized states which describes both initial and final states of the decay up to second order. We use the new scheme to construct the Fano-ADC(2,2) approximation for the decay widths and show that it provides superior accuracy for the decay widths of a series of processes. Moreover, the Fano-ADC(2,2) method provides access to second-order decay processes, such as double Auger decay, which are qualitatively beyond reach of the previously available Fano-ADC implementations.     
\end{abstract}


\maketitle 

\section{Introduction}
\label{sec:Intro}

Inner-shell excitation or ionization of an atom or a molecule produces metastable electronic states which can decay by  autoionization. The most common such relaxation process is Auger decay\cite{Auger1925jpr} (AD), in which the vacancy is refilled by a valence electron and other electron is emitted to continuum. Closely related process is the interatomic Coulombic decay\cite{Cederbaum1997prl,Jahnke2015jpb} (ICD), involving energy transfer from the ionized or excited species to its environment and emission of secondary electron from a neighboring atom or molecule. Less frequently, electron correlation leads to higher-order multi-electron transitions. Basic examples comprise simultaneous emission of two electrons in the double Auger decay\cite{Carlson1965prl,Journel2008pra,Roos2018scirep} (DAD) or double ICD (DICD)\cite{Averbukh2006prl} processes, or collective recombination of two or more vacancies with emission of a single electron\cite{Feifel2016prl,Zitnik2016pra,Averbukh2009prl}. Interest in these higher-order processes intensified recently due to their significance in multiply ionized or excited systems produced, e.g., after irradiation by high-intensity free-electron lasers.

Among fundamental characteristics of a metastable state (resonance) belongs its decay width, the knowledge of which is essential for understanding of the dynamics of the system following the excitation. Although wide range of theoretical approaches is available for its computation, transitions with continuum electrons are inherently more complicated to describe than processes involving only bound states. Number of studies of AD employ many-body perturbation theory\cite{Kelly1975pra,Amusia1992pra} or various many-electron wave function models combined with the golden rule formula for transition probabilities\cite{Pindzola1987pra,Chen1991pra}. To account for the higher-order multi-electron transitions, approximate formulas corresponding to different decay mechanisms such as shake-off or knock-out are often used\cite{Zeng2013pra}.

Distinct class of \emph{ab initio} methods builds on the highly developed computational quantum chemistry for bound states and describe the decay processes using square integrable ($\lsq$) basis sets. One possible approach is to introduce complex absorbing potential\cite{Riss1993jpb} (CAP) into the electronic Hamiltonian in order to transform the divergent Siegert states associated with resonances into $\lsq$ wave functions. In this sense, CAP approach is closely related to the exterior complex scaling transformation\cite{Moiseyev2011book}. The resulting non-Hermitian Hamiltonian can then be represented employing various quantum chemical models for excited states, such as configuration interaction\chng{\cite{Santra2002prep,Peng2016jcp,Zhang2012pra}}, algebraic diagrammatic construction\cite{Schirmer1991pra,Vaval2007jcp} (ADC), or equation-of-motion coupled-cluster\cite{Ghosh2013pccp}.

Another family of $\lsq$ \emph{ab initio} methods for computation of intra- and inter-atomic non-radiative decay widths stemming from the Fano-Feshbach theory of resonances  are the Fano-ADC techniques\cite{Averbukh2005jcp,Gokhberg2007jcp,Kolorenc2008jcp}. They rely on the Fano theory of resonances\cite{Fano1961pr,Howat1978jpb}, ADC in the intermediate state representation\cite{Mertins1996pra,SchirmerMBM2018} (ISR) for the many-electron wave functions, and Stieltjes imaging technique\cite{Langhoff1979empmc} to recover correct normalization of the discretized $\lsq$ representation of the continuum. Over the last decade, Fano-ADC was established as possibly the most efficient approach for calculations of intra- and interatomic decay widths. It has been utilized in a series of studies of ICD\cite{Sisourat2010nphys,Gokhberg2014nat} and made possible prediction of new collective three-electron decay processes\cite{Averbukh2009prl,Feifel2016prl}. Among the   attractive properties of the method are size consistency of the ADC scheme, capability of producing converged decay widths over many orders of magnitude, and the possibility to estimate the partial decay widths despite the lack of proper continuum wave functions\cite{Averbukh2005jcp}.
The high efficiency of the method is connected with the so-called compactness of the ADC expansion, which involves the smallest possible explicit configuration space needed at a given order of perturbation theory and lack of need to diagonalize the full \chng{ADC(n) Hamiltonian matrix} (as is required in the complex scaling and CAP-based techniques), but rather only the restricted Hamiltonian represented \chng{in the subsets of configurations corresponding to initial and final state subspaces, respectively.}

With the emerging time-domain spectroscopy techniques with attosecond resolution\cite{Drescher2002nat,Huetten2018ncomm} and the increasing interest in multiple-ionization processes, the demands on the accuracy and scope of the theoretical methods increases rapidly. At present, the available implementations of the Fano-ADC method are limited to the extended second order ADC schemes, ADC(2)x. For singly ionized systems, ADC(2)x comprises only the two lowest excitation classes, which can be characterized (with respect to the neutral ground state configuration) as one-hole ($1h$) and two-hole-one-particle ($2h1p$). At this level, the expansion of the correlated $1h$-like main ionic states is complete through the second order of perturbation theory (PT) while that of the $2h1p$-like ionization satellite states only through the first order. Since typical AD or ICD process corresponds to a transition from $1h$-like inner-shell vacancy state to a $2h1p$-like state, representing the resulting doubly ionized system plus an electron in continuum, such correlation imbalance can affect the accuracy of the computation of the relevant coupling matrix elements. Furthermore, first order PT is often simply insufficient to describe the correlation in the satellite states adequately, severely limiting the usability of the method for studying decay of the $2h1p$-like ionized-excited states\cite{Rist2017chp}. Finally, the processes involving emission of two electrons into continuum, such as DAD and DICD, are qualitatively not accessible within ADC(2)x approximations because of the absence of the main excitation classes of the corresponding final states of the decay, e.g.\ 3h2p for DAD of single core vacancies.

To remedy these issues, we have exploited the flexibility of the ISR approach to develop ISR-ADC(2,2) scheme in which the satellite states are treated consistently with the main ionic states through the second order of PT. The resulting expansion involves also the $3h2p$ excitation class, opening the possibility to study decay processes accompanied by emission of two electrons (DAD, DICD). In this work, we present the  Fano-ADC(2,2) method and test its capabilities to produce accurate decay widths on various examples of intra- and interatomic decay processes, with emphasis on the multi-electron transitions. The paper is organized as follows: in Sec.\ \ref{sec:ADCmain}, we give an overview of the ADC procedure, the intermediate state representation, and infer the structure of the ISR-ADC(2,2) scheme. In Sec.\ \ref{sec:FanoADC22} we summarize the Fano theory of resonances and describe in detail the Fano-ADC methodology. In Sec.\ \ref{sec:IPs}, we demonstrate the balanced description of the main and satellite ionization states by computing atomic ionization potentials. Results of the decay widths calculations are given and discussed in Secs.\ \ref{sec:Auger}--\ref{sec:ICD}, and the paper is concluded in Sec.\ \ref{sec:conclusions}.

\section{Algebraic Diagrammatic Construction for electron propagator}
\label{sec:ADCmain}
In energy representation, the electron propagator reads
\begin{eqnarray}
 \label{eq:elprop}
 G_{pq}(\omega) & = & \langle\Psi_0|c_p\left(\omega-H+E_0+i\eta\right)^{-1}c^\dag_q|\Psi_0\rangle + \langle\Psi_0|c^\dag_q\left(\omega+H-E_0-i\eta\right)^{-1}c_p|\Psi_0\rangle \nonumber \\
  & = & G^+_{pq}(\omega)+G^-_{pq}(\omega).
\end{eqnarray}
Here, $H$ is the Hamiltonian operator, $|\Psi_0\rangle$ is the exact $N$-electron ground state wave function and $E_0$ the ground state energy. $c^\dag_p$ and $c_p$ are electron creation and annihilation operators associated with a basis of one-particle states $|p\rangle$ -- usually Hartree-Fock (HF) orbitals -- and $\eta$ is positive infinitesimal convergence factor. We will focus in particular on the $(N-1)$-electron part $G^-_{pq}(\omega)$ which is relevant for the description of the ionization process. It can be cast into the Lehmann spectral representation,
\begin{equation}
 \label{eq:Lehmann}
 G^-_{pq}(\omega) = \sum_n\frac{\langle\Psi_0|c^\dag_q|\Psi_n^{N-1}\rangle\langle\Psi_n^{N-1}|c_p|\Psi_0\rangle}{\omega+E_n^{N-1}-E_0-i\eta},
\end{equation}
via introduction of complete set of exact $(N-1)$-electron states $|\Psi_n^{N-1}\rangle$. In this representation, the electron propagator is given as a sum of simple poles located at the negative ionization energies,
\begin{equation}
 \label{eq:ione}
 -I_n = E_0-E_n^{N-1}.
\end{equation}


Direct ADC procedure \cite{Schirmer1982pra,Schirmer1983pra} is an approach which enables to systematically derive hierarchy of approximations ADC($n$) to the electron propagator (or other type of many-electron Green's function), which are complete up to order $n$ of PT and include infinite partial summations needed to recover the characteristic simple poles structure of the propagator, highlighted in the Lehmann representation. First, closed-form algebraic ansatz is imposed on the matrix $\mathbf{G}^-$,
\begin{equation}
 \label{eq:ADCansatz}
 \mathbf{G}^-(\omega)=\mathbf{f}^\dag(\omega-\mathbf{K}-\mathbf{C})^{-1}\mathbf{f},
\end{equation}
where the \emph{ADC secular matrix} $\mathbf{K}+\mathbf{C}$ consists of the diagonal matrix $\mathbf{K}$ of zero-order ionization energies and the hermitian \emph{effective interaction matrix} $\mathbf{C}$. $\mathbf{f}$ are \emph{effective transition amplitudes}. The infinitesimal $-i\eta$ is not essential in the following and will be omitted. The ADC form \eqref{eq:ADCansatz} can be expanded in a formal perturbation series, assuming the existence of perturbation expansion of the ADC secular matrix and transition amplitudes,
\begin{eqnarray}
 \label{eq:PTforC}
 \mathbf{K}+\mathbf{C}= && \mathbf{K}^{(0)}+\mathbf{C}^{(1)}+\mathbf{C}^{(2)}+\dots \\
 \label{eq:PTforf}
 \mathbf{f} = && \mathbf{f}^{(0)}+\mathbf{f}^{(1)}+\mathbf{f}^{(2)}+\dots
\end{eqnarray}
The ADC($n$) approximation scheme (i.e., explicit expressions for $\mathbf{K}$, $\mathbf{C}^{(i)}$ and $\mathbf{f}^{(i)}$) is then obtained through comparison of the formal perturbation expansion of Eq.\ \eqref{eq:ADCansatz} with the standard diagrammatic perturbation expansion for $\mathbf{G}^-$, truncated after the PT order $n$.

Once the (exact or approximate) ADC secular matrix is available, its diagonalization
\begin{equation}
 \label{eq:ADCsecular}
 (\mathbf{K}+\mathbf{C})\mathbf{X}=\mathbf{X}\mathbf{\Omega},\qquad \mathbf{X}^T\mathbf{X}=\mathbf{1}
\end{equation}
provides the physical information contained in the propagator $\mathbf{G}^-$. In particular, $\mathbf{\Omega}$ is the diagonal matrix of eigenvalues $\omega_n$ which correspond to negative ionization energies $-I_n$. An important characteristic of the ADC approach is that the secular problem $\eqref{eq:ADCsecular}$ is Hermitian. Furthermore, it can be shown that its PT expansion is regular, i.e., the energy denominators appearing in the expansion of the elements of $\mathbf{C}$ are larger than the energy gap between occupied and virtual HF orbitals. For a detailed and pedagogical account on ADC and related methods we refer the reader to the book of J.\ Schirmer\cite{SchirmerMBM2018}.

\subsection{Intermediate state representation}
\label{sec:ISR}
As an alternative to the diagrammatic derivation, ADC can be formulated in the so-called intermediate state representation (ISR) \cite{Schirmer1991pra,Mertins1996pra}. ISR-ADC provides closed-form version of the ADC secular problem \eqref{eq:ADCsecular}, which is equivalent to the direct procedure but is in fact a wave-function method, which significantly extends its scope of applicability. In particular, the availability of explicit representation of the $(N-1)$-electron states $|\Psi_n^{N-1}\rangle$ makes it possible to evaluate matrix elements of general operators\cite{Schirmer2004jcp} or to compute coupling matrix elements driving the decay of metastable states. Furthermore, the ISR approach is more flexible in construction of the perturbation expansion of the effective interaction matrix $\mathbf{C}$, which allows us to devise the desired ISR-ADC(2,2) scheme with balanced representation of main and satellite ionization states.

The ISR-ADC approach is based on the observation that the non-diagonal ADC representation \eqref{eq:ADCansatz} of the electron propagator can be obtained from the general formula \eqref{eq:elprop} by using a complete basis of some $(N-1)$-electron intermediate states $\isr{k}$ (IS) instead of the exact Hamiltonian eigenstates $|\Psi_n^{N-1}\rangle$. This suggests that the ADC secular matrix $\mathbf{M}=-(\mathbf{K}+\mathbf{C})$ can be interpreted as a representation of the shifted Hamiltonian $H-E_0$ in a basis of appropriate ISs.

The particular set of ISs leading to representation of the electron propagator equivalent to the direct ADC approach described above can be constructed explicitly without a reference to diagrammatic PT \cite{Mertins1996pra,SchirmerMBM2018}. The procedure starts by introducing so-called \emph{correlated excited states} (CES)
\begin{equation}
 \label{eq:CES}
 \CE{J} = C_J|\Psi_0\rangle,
\end{equation}
where $C_J$ are the physical excitation operators associated with one-particle basis of HF orbitals,
\begin{equation}
 \label{eq:CJ}
 \{C_J\} = \left\{c_k; c^\dag_a c_k c_l,k<l;c_a^\dag c_b^\dag c_j c_k c_l,a<b,j<k<l;\dots\right\}.
\end{equation}
The indices $j,k,l,\dots$ and $a,b,\dots$ correspond to occupied and virtual HF orbitals, respectively. The CESs can therefore be classified into excitation classes as $1h$, $2h1p$, $3h2p$, and so on. In the following, $J$ corresponds to individual configuration while $[J]=\mu$ denotes the whole $\mu h-(\mu-1)p$ class.

Unlike the configurations
\begin{equation}
 \label{eq:HFconf}
 |\Phi_J\rangle = C_J|\Phi_0\rangle
\end{equation}
derived from the HF ground state $|\Phi_0\rangle$, which form basis of the standard Configuration Interaction (CI) expansion, CESs are not orthonormal.
\chng{The non-orthogonality stems from perturbation corrections to the ground state wave function. For instance, second order correction brings about admixture of $1h1p$ excitations into  $|\Psi_0\rangle$\cite{SchirmerMBM2018}, which upon action of $c_k$ translates into $2h1p$ contributions to the $1h$ CESs. In turn, CESs from $1h$ and $2h1p$ excitation classes are no longer orthogonal.}
It is the specific \emph{excitation class orthogonalization} (ECO) procedure which leads to the ISR-ADC representation. ECO proceeds iteratively as follows. Assuming ISs $\isr{K}$ belonging to excitation classes $[K]=1,\dots,\nu-1$ are available, \emph{precursor states} $|\Psi_J^\#\rangle$ belonging to the class $[J]=\nu$ are constructed through Gram-Schmidt orthogonalization of the CESs $|\Psi_J^0\rangle$ with respect to the ISs belonging to all lower excitation classes as
\begin{equation}
 \label{eq:precursor}
 |\Psi_J^\#\rangle = \CE{J}-\sum_{[K]<[J]}\isr{K}\langle\tilde{\Psi}_K\CE{J}.
\end{equation}
The ISs of the class $[J]=\nu$ are then obtained via symmetric orthogonalization of the precursor states within the excitation class,
\begin{equation}
 \label{eq:ISRsymOG}
 \isr{J} = \sum_{[I]=\nu}|\Psi^\#_I\rangle(\mathbf{S}_\nu^{-1/2})_{IJ}.
\end{equation}
Here, $\mathbf{S}_\nu$ is the overlap matrix
\begin{equation}
 \label{eq:precursorS}
 (\mathbf{S}_\nu)_{IJ} = \langle\Psi_I^\#|\Psi_J^\#\rangle
\end{equation}
of the precursor states belonging to the excitation class $[I]=[J]=\nu$. Note that the procedure can be initiated correctly starting from the lowest $[J]=1$ class as the precursor states $|\Psi^\#_{[J]=1}\rangle$ are equal to CESs $|\Psi_{[J]=1}^0\rangle$ and the Gram-Schmidt orthogonalization step \eqref{eq:precursor} does not apply.

Starting from the exact ground state $|\Psi_0\rangle$ and complete manifold of excitation operators~\eqref{eq:CJ}, ECO procedure leads to exact representation of the shifted Hamiltonian (secular matrix),
\begin{equation}
 \label{eq:MIJ}
 M_{IJ} = \langle\tilde{\Psi}_I|H-E_0\isr{J}.
\end{equation}
Practical computation scheme is obtained by using truncated PT expansion for the ground state (note the \emph{intermediate normalization} $\langle\Phi_0|\Psi_0\rangle=1$),
\begin{equation}
 \label{eq:GSPTn}
 |\Psi_0\rangle = |\Phi_0\rangle + |\Psi_0^{(1)}\rangle + \cdots +|\Psi_0^{(n)}\rangle + O(n+1).
\end{equation}
This in turn leads naturally to PT expansion of the ISs, 
\begin{equation}
 \label{eq:ISpt}
 \isr{J} = |\Phi_J\rangle+|\tilde{\Psi}_J^{(1)}\rangle+\cdots+|\tilde{\Psi}_J^{(n)}\rangle+O(n+1),
\end{equation}
and, together with the standard PT expansion for the ground state energy,
\begin{equation}
 \label{eq:E0pt}
 E_0 = E_0^{(0)}+E_0^{(1)}+\dots+E_0^{(n)}+O(n+1),
\end{equation}
to expansion of the secular matrix elements
\begin{equation}
 \label{eq:MIJpt}
 M_{IJ} = M_{IJ}^{(0)}+M_{IJ}^{(1)}+\cdots+M_{IJ}^{(n)} + O(n+1).
\end{equation}
Corresponding PT expansion of the transition amplitudes $\mathbf{f}$ can be derived in the same manner but is redundant for our purposes.

An $n$-th order ISR-ADC approximation equivalent to ADC($n$) derived by the direct procedure is obtained by truncating the expansion \eqref{eq:MIJpt} for each $IJ$ block at an appropriate order. Fundamental example is the second order ADC($2$) scheme. The explicit configuration space is spanned by $1h$ and $2h1p$ ISs and the secular matrix has the block structure\cite{SchirmerMBM2018}
\begin{eqnarray}
 \label{eq:ADC2blocks}
 M_{1h,1h} = && M_{1h,1h}^{(0)} + M_{1h,1h}^{(2)} \nonumber \\
 M_{1h,2h1p} = && M_{1h,2h1p}^{(1)} \\
 M_{2h1p,2h1p} = && M_{2h1p,2h1p}^{(0)}.\nonumber
\end{eqnarray}
Together with corresponding approximation of the effective transition amplitudes $\mathbf{f}$, this scheme provides complete second order representation of $\mathbf{G}^-$. However, description of correlated $1h$- and $2h1p$-like states is inconsistent, as will be shown in the following subsection.

\subsection{Canonical order relations}
The equivalence of the direct ADC and ISR-ADC formulations rests on two common features\cite{SchirmerMBM2018}. One is the separability of the secular matrix with respect to non-interacting subsystems, which leads to size-consistency of the approximation at any given order. The other are the \emph{canonical order relations} (COR) fulfilled by the secular matrix $\mathbf{M}$: the PT expansions \eqref{eq:MIJpt} of the off-diagonal ($[I]\neq[J]$) matrix elements do not begin at the zeroth order but rather follow the general rule\cite{Mertins1996pra,SchirmerMBM2018}
\begin{equation}
 \label{eq:COR}
 M_{IJ} \sim O(|[I]-[J]|).
\end{equation}
For instance, the lowest order contribution to matrix elements $M_{1h,3h2p}$, which couple the $1h$ and $3h2p$ excitation classes, are of the second order, $M_{1h,3h2p}\sim O(2)$, see 
Table~12.1 of Ref.~\onlinecite{SchirmerMBM2018} for further details.

From the COR \eqref{eq:COR} it is possible to determine the PT order of an error of ionization energies computed using any given truncated ADC scheme. Let us assume that an eigenvector $X_I$ of the secular matrix can be classified as belonging to the excitation class $\nu$, that is, its expansion is dominated by class $\nu$ ISs. Corresponding eigenvalue, $\omega_I$, is then linear in the matrix elements $M_{IJ}$ belonging to the respective diagonal block of the secular matrix ($[I]=[J]=\nu$) and quadratic in the matrix elements $M_{IK}$ belonging to the off-diagonal blocks directly coupling classes $[I]=\nu$ and $[K]\ne\nu$. The PT order of the error of the eigenvalue is then given by the lowest order correction missing in the secular matrix.

As an example, consider the ADC(2) scheme \eqref{eq:ADC2blocks}. For a $1h$ class, the critical correction missing is the third-order $M_{1h,1h}^{(3)}$, therefore, corresponding eigenenergies are correct through the second order of PT. As noted above, coupling to the absent $3h2p$ excitation class is of the second order and would contribute to the $1h$ state energies only by a fourth-order correction, together with the neglected second-order $M_{1h,2h1p}^{(2)}$ coupling to the $2h1p$ class. In contrast, energies of the $2h1p$ states are complete only through the zeroth order. Adding first order elements $M_{2h1p,2h1p}^{(1)}$ to the $2h1p/2h1p$ block leads to the so-called extended second-order [ADC(2)x] scheme and improves the $2h1p$-state energies by one order of PT.

COR lie behind the \emph{compactness} property\cite{Mertins1996pra} of the ADC secular matrix. It is best demonstrated in comparison with the CI method, in which the $|\Psi_n^{N-1}\rangle$ states are expanded in terms of the HF configurations \eqref{eq:HFconf}. In the resulting matrix representation of the $(N-1)$-electron Hamiltonian, $\mathbf{H}$, each excitation class $\mu$ is coupled with up to four adjacent classes, $\mu\pm 1$ and $\mu\pm 2$, by first order matrix elements. In particular, $H_{1h,3h2p}\sim O(1)$ and the $3h2p$ excitation class has to be included explicitly into the configuration space in order to recover all second order contributions to the $1h$-state ionization energies. At the ADC(2) level, on the other hand, first order coupling between $1h$ and $3h2p$ classes is taken into account implicitly through the second order $M_{1h,1h}^{(2)}$ matrix elements, which leads to significant reduction of the computational demands.

\chng{
The ISR concept and related compactness property are not unique to the ADC methodology. Within the coupled-cluster (CC) framework, treatment of excited or ionized states is based on CESs derived from the CC ground state parametrization. It leads to the so-called biorthogonal coupled-cluster (BCC) representation in which the Hamiltonian is given by a non-hermitian secular matrix. Due to the different quality of the two underlying (left and right) biorthogonal sets of states, the BCC secular matrix combine COR and CI-type order structure. The errors of excitation or ionization energies computed using BCC schemes thus lie between the CI and ADC expansions truncated after the same excitation class. Detailed analysis of the relation between ISR-ADC and BCC can be found in the literature.\cite{Mertins1996pra,Schirmer2009tca}
}

\subsection{ISR-ADC(2,2) approximation scheme}
\label{sec:ADC22}
From the above discussion of COR, it is now straightforward to design an ISR-ADC(2,2) scheme in which both the $1h$ and $2h1p$-state energies will be determined consistently through the second order of PT. Starting from the ISR-ADC(2) of Eq.\ \eqref{eq:ADC2blocks}, it follows that the ISR-ADC(2,2) secular matrix has to include also the first- and second order matrix elements $M_{2h1p,2h1p}^{(1)}$, $M_{2h1p,2h1p}^{(2)}$ in the diagonal $2h1p/2h1p$ block. Furthermore, the first order $M_{2h1p,3h2p}^{(1)}$ matrix elements directly coupling the $2h1p$ and $3h2p$ ISs are needed as they contribute to the $2h1p$ states energies by second order correction. Therefore, $3h2p$ excitation class has to be included in the explicit configuration space. Corresponding diagonal block of the secular matrix must contain at least zero order contribution $M_{3h2p,3h2p}^{(0)}$, which again contribute to the $2h1p$ energies at second order.

This structure of the secular matrix constitutes the minimal scheme fulfilling the requirements. However, to extend the applicability of the method, it is desirable to improve the description of the $3h2p$ states to at least first order of PT. This is achieved by including first order matrix elements to the $3h2p/3h2p$ block. These terms contribute to the $2h1p$ state energies at third order through corrections of the type
\begin{equation}
    \label{eq:3h2p_to_2h1p}
    \omega_{2h1p}\propto (M^{(1)}_{2h1p,3h2p})^2 M^{(1)}_{3h2p,3h2p}.
\end{equation}
Corresponding correction to the $1h$ state energies is only of fifth order,
\begin{equation}
    \label{eq:3h2p_to_1h}
    \omega_{1h}\propto (M^{(1)}_{1h,2h1p})^2 (M^{(1)}_{2h1p,3h2p})^2 M^{(1)}_{3h2p,3h2p}.
\end{equation}
This extension significantly improves the consistency between $1h$ and $2h1p$-state energies since the correction \eqref{eq:3h2p_to_2h1p} compensates the respective third order contribution of $M^{(1)}_{2h1p,2h1p}$ to the $1h$ state energies,
\begin{equation}
    \label{eq:2h1p_to_1h}
    \omega_{1h}\propto (M^{(1)}_{1h,2h1p})^2 M^{(1)}_{2h1p,2h1p}.
\end{equation}
Furthermore, we will show in Sec.\ \ref{sec:Auger} that inclusion of the second order contributions $M_{1h,2h1p}^{(2)}$ to the $1h/2h1p$ coupling matrix elements is necessary to obtain accurate Auger decay widths. Concerning the eigenenergies, these terms contribute to both the $1h$ and $2h1p$-state energies at third order,
\begin{equation}
    \label{eq:1h2h1p2_to_1h,2h1p}
    \omega_{1h},\omega_{2h1p} \propto M^{(1)}_{1h,2h1p} M^{(2)}_{1h,2h1p}.
\end{equation}
However, as both the $1h$ and $2h1p$ ISs have to be expanded through second order to arrive at the ISR-ADC(2,2) approximation, is appears to be vital to account for the direct coupling at the same level.

Thorough the rest of the paper, we will drop the ISR prefix an denote the minimal scheme as \adcm{}, scheme extended by the first-order $3h2p/3h2p$ matrix elements as \adcx{}, and the full scheme including also the second order $1h/2h1p$ couplings as \adcf{}. For clarity, block structure of different variants of the proposed ADC(2,2) scheme, together with commonly used ADC($n$) schemes from the standard ADC hierarchy, is summarized in Tab.\ \ref{tab:ADC22secular}.
\chng{
Computational cost of \adcf{} and \adcx{} scales with the system size as $n_{occ}^3n_{virt}^4$, $n_{occ}$ and $n_{virt}$ being the numbers of occupied and virtual molecular orbitals, respectively. This scaling is determined by the number of nonzero first-order $3h2p/3h2p$ matrix elements
and thus applies both to the matrix construction and matrix-vector multiplication. The latter determines the cost of iterative diagonalization methods. \adcm{} scales more favorably as $n_{occ}^4 n_{virt}^3$ for matrix construction and as $n_{occ}^4 n_{virt}^2$ for matrix-vector multiplication. For comparison, the scaling of the commonly used ADC(2)x is $n_{occ}^3 n_{virt}^2$.
}
\begin{table}[htb]
 \renewcommand{\arraystretch}{1.6}
 \begin{tabular}{c|C{4em}|C{4em}|C{4em}|}
   \mc{}& \mc{$1h$} & \mc{$2h1p$} & \mc{$3h2p$} \\
  \cline{2-4}
  $1h$ & $M_{11}^{(\mu)}$ & $M_{12}^{(\nu)}$ & -- \\
  \cline{2-4}
  $2h1p$ & $M_{21}^{(\nu)}$ & $M_{22}^{(\xi)}$ & $M_{23}^{(\sigma)}$ \\
  \cline{2-4}
  $3h2p$ & --  & $M_{32}^{(\sigma)}$ & $M_{33}^{(\chi)}$ \\
  \cline{2-4}
 \end{tabular}\\[3ex]
 \renewcommand{\arraystretch}{1.0}
 \begin{tabular}{|c||C{3em}|C{3em}|C{3em}|C{3em}|C{3em}|}
  \hline
  scheme & $M_{11}^{(\mu)}$ & $M_{12}^{(\nu)}$ & $M_{22}^{(\xi)}$ & $M_{23}^{(\sigma)}$ & $M_{33}^{(\chi)}$ \\
  \hline
  \hline
  ADC(2) & 0,2 & 1 & 0 & - & - \\
  \hline
  ADC(2)x & 0,2 & 1 & 0,1 & - & - \\
  \hline
  ADC(3) & 0,2,3 & 1,2 & 0,1 & - & - \\
  \hline
  \adcm & 0,2 & 1 & 0,1,2 & 1 & 0 \\
  \hline
  \adcx & 0,2 & 1 & 0,1,2 & 1 & 0,1 \\
  \hline
  \adcf & 0,2 & 1,2 & 0,1,2 & 1 & 0,1\\
  \hline
\end{tabular}
 \caption{\label{tab:ADC22secular}Block structure (perturbation theory orders) of the secular matrix $\mathbf{M}$ in standard ADC($n$) schemes and the three proposed variants of ADC(2,2).}
\end{table}

So far, we have only considered the PT expansion of the eigenvalues of the ADC(2,2) secular matrix. Before concluding this section, comment on the resulting representation of the $(N-1)$-electron wave functions $|\Psi^{N-1}_n\rangle$ is in order. In the ISR-ADC scheme, these are given in terms of ISs as the eigenvectors of the secular matrix, i.e., by the matrix $\mathbf{X}$ in Eq.\ \eqref{eq:ADCsecular}. Assuming $|\Psi^{N-1}_n\rangle$ can be classified as belonging to the excitation class $[n]$, the COR structure \eqref{eq:COR} of the secular matrix $\mathbf{M}$ implies order relations for the eigenvectors in the form\cite{SchirmerMBM2018}
\begin{equation}
 \label{eq:CORX}
 X_{Jn}=\langle\tilde{\Psi}_J|\Psi^{N-1}_n\rangle\sim O(|[J]-[n]|).
\end{equation}
It follows that in the \adcf{} scheme given in Tab.\ \ref{tab:ADC22secular}, $|\Psi^{N-1}_n\rangle$ belonging to both $1h$ and $2h1p$ excitation classes are represented fully only through the first order of PT. For the $1h$ states, the error is determined by the neglected second order direct coupling $M_{1h,3h2p}^{(2)}$ to the $3h2p$ class. For the $2h1p$ states, $M_{2h1p,3h2p}^{(2)}$ and $M_{2h1p,4h3p}^{(2)}$ matrix elements are necessary to account for all second order contributions to the respective wave functions. Inclusion of the $M_{1h,3h2p}^{(2)}$ matrix elements would in principle be possible. However, explicit appearance of the $4h3p$ excitation class would result to an impractical method, computationally intractable even for the smallest atomic systems.

\section{Fano-ADC(2,2) method}
\label{sec:FanoADC22}

In this section, we describe the Fano-ADC methodology for computation of non-radiative decay rates and point out the necessary modification of the previous implementations, related to the use of ADC(2,2) scheme. We start by reviewing the Fano theory of resonances, followed by description how the theory can be combined with ISR-ADC representation of the many-electron wave functions employing an $\lsq$ one-electron basis set.

\subsection{Fano theory of resonances}
\label{sec:Fano}

In the theory of resonances developed by Fano\cite{Fano1961pr} and formulated in a convenient pro\-jec\-ti\-on-operator formalism by Feshbach\cite{Feshbach1964rmp}, the solution $|\Psi_{E,\alpha}\rangle$ of the time-independent Schr\"odinger equation (TISE)
 \begin{equation}
   \label{eq:TISE}
   H|\Psi_{E,\alpha}\rangle = E|\Psi_{E,\alpha}\rangle
 \end{equation}
 at some (real) energy $E$ near resonance is represented as a superposition of bound state-like $\lsq$ discrete state $|\Phi\rangle$ and background continuum components $|\chi_{\beta,\epsilon}\rangle$,
 \begin{equation}
  \label{eq:Fano_ansatz}
  |\Psi_{E,\alpha}\rangle = a_\alpha(E)|\Phi\rangle+\sum_{\beta=1}^{N_c}\int C_{\beta,\alpha}(E,\epsilon)|\chi_{\beta,\epsilon}\rangle d\epsilon.
 \end{equation}
Here, $H$ is the full electronic Hamiltonian, $N_c$ is the number of available decay channels, index $\alpha=1\,\dots,N_c$ numbers the independent solutions and $\epsilon$ is the energy of emitted particle.

The decomposition \eqref{eq:Fano_ansatz} corresponds to partitioning of the Hilbert space into the continuum subspace $\Psub$ and the subspace $\Qsub$ which contains the bound-like discrete state. It can be realized through introduction of the corresponding projection operators,
 \begin{equation}
  \label{QPprojectors}
  Q= |\Phi\rangle\langle\Phi|\quad \text{and}\quad P=\sum_{\beta=1}^{N_c}\int|\chi_{\beta,\epsilon}\rangle\langle\chi_{\beta,\epsilon}| d\epsilon.
 \end{equation}
Typically, the two projectors are constructed as complementary, $P=\mathds{1} -Q$. However, an important feature of the Fano theory is that orthogonality between the two projectors is not strictly required. In some applications, particularly in connection with quantum-chemical calculations carried out in $\lsq$ basis set, non-orthogonal bound-like and continuum-like subspaces arise naturally \cite{Yun2018jcp}.

When applied to a decay process, the discrete state $|\Phi\rangle$ and the background continuum states $|\chi_{\beta,\epsilon}\rangle$ can be associated with the initial and final states, respectively. The discrete state is characterized by its mean energy
\begin{equation}
 \label{eq:Ed}
 E_\Phi=\langle\Phi|H|\Phi\rangle
\end{equation}
and the background continuum functions $|\chi_{\beta,\epsilon}\rangle$ are assumed to diagonalize the Hamiltonian to a good approximation,
\begin{equation}
 \label{eq:bg_continuum}
 \langle\chi_{\beta',\epsilon'}|H-E|\chi_{\beta,\epsilon}\rangle \approx (E_\beta+\epsilon-E)\delta_{\beta',\beta}\delta(E_{\beta'}+\epsilon'-E_\beta-\epsilon).
\end{equation}
Here, $E_\beta$ is energy of the resulting $(N-1)$-electron (with respect to $|\Phi\rangle$) ionic state and the $N_c$ open decay channels in Eq.\ \eqref{eq:Fano_ansatz} then correspond to the energetically accessible states with energies $E_\beta<E_\Phi$. Defining the width function\cite{Howat1978jpb} as a sum of partial widths,
\begin{equation}
 \label{eq:GammaE}
 \Gamma(E) =  \sum_{\beta=1}^{N_c}\Gamma_\beta(E)=2\pi\sum_{\beta=1}^{N_c}\left|\langle\Phi|H-E|\chi_{\beta,E-E_\beta}\rangle\right|^2,
\end{equation}
and solving the TISE \eqref{eq:TISE}, the coefficient $a_\alpha(E)$ can be expressed as a generalized Lorentzian,
\begin{equation}
 \label{eq:aalpha}
 |a_\alpha(E)|^2=\frac{1}{2\pi}\frac{\Gamma_\alpha(E)}{(E-E_r)^2+\Gamma(E)^2/4},
\end{equation}
with the energy
\begin{equation}
 \label{eq:Eres}
 E_r = E_\Phi+\Delta(E) = E_\Phi+\sum_{\beta=1}^{N_c}\! \mathscr{P}\!\!\int\frac{|\langle\Phi|H-E|\chi_{\beta,\epsilon}\rangle|^2}{E-E_\beta-\epsilon}d\epsilon
\end{equation}
defining position of the resonance and $\Gamma=\Gamma(E_r)$ its width ($\mathscr{P}$ stands for principal value integration). In practical applications, in particular those involving $\lsq$ basis, only discretized approximation of the width function $\Gamma(E)$ is acquired and evaluation of the level shift function $\Delta(E)$ is not feasible. Therefore, the resonance energy and width are usually approximated as $E_r\approx E_\Phi$ and $\Gamma\approx\Gamma(E_\Phi)$.

In the case of strong channel mixing, Eq.\ \eqref{eq:bg_continuum} is not satisfied for the continuum states $|\chi_{\beta,\epsilon}\rangle$ associated with the decay channels. In such a case, unitary transformation (prediagonalization) of the background continuum is necessary,
\begin{equation}
 \label{eq:chi_prediag}
 |\chi^-_{\lambda,\epsilon}\rangle = \sum_{\beta=1}^{N_c}\int D_{\lambda,\beta}(\epsilon,\epsilon')|\chi_{\beta,\epsilon'}\rangle d\epsilon'.
\end{equation}
The rest of the procedure is completely analogical including the formula for the total width, only the expression for partial widths associated with the original channels $\beta$ becomes more involved \cite{Howat1978jpb}.

\subsection{Fano theory in the framework of ISR-ADC}
\label{sec:FanoISR}

Since we are interested in decay of a single inner shell vacancy states, the wave function $|\Psi_{E,\alpha}\rangle$ \eqref{eq:Fano_ansatz} is an $(N-1)$-electron wave function. In this section we describe how the bound-like discrete component $|\Phi\rangle$ and the continuum components $|\chi_{\beta,\epsilon}\rangle$ can be approximated in the framework of an ISR-ADC scheme implemented using an $\lsq$ basis. To this end, we need to divide the configuration space spanned by the ISs into the continuum subspace $\Psub$, containing the final states of the decay, and the subspace $\Qsub$ containing the bound states and bound-like discrete components associated with the metastable states.

In a rigorous theory, the fundamental difference between the two subspaces is that the $\Psub$ subspace contains states with at least one electron in continuum while $\Qsub$ contains strictly $\lsq$ wave functions. In any method employing $\lsq$ one-particle basis sets, however, this distinction is lost as the continuum is discretized and approximated by $\lsq$ wave functions. Therefore, other criteria of the classification of ISs have to be devised. The essential requirement is that, within the $\Qsub$ subspace itself, the discrete component is bound and can only decay through coupling to the background continuum of the $\Psub$ subspace. Therefore, any representation of $|\Phi\rangle$ must not contain any contribution corresponding to the possible final states of the decay. Within the ADC(2,2) configuration space, ISs corresponding to the final states of decay of singly ionized states are to be found in the $2h1p$ and $3h2p$ excitation classes as the continuum electron has to be described by a virtual orbital. To identify specific ISs belonging to the $\Psub$ subspace, various schemes applicable in different situations are possible.

The simplest approach, based on the lowest-order estimates for the energies $E_\beta$ of $2h$ ($[\beta]=2$) and $3h$ ($[\beta]=3$) decay channels is described in the original work of Averbukh and Cederbaum\cite{Averbukh2005jcp}. Due to its rather limited applicability, this approach was not implemented in the present work. Instead, we employ two different schemes, one based on hole localization\cite{Averbukh2005jcp} and the other on the adaptation (prediagonalization) of ISs\cite{Kolorenc2015jcp}.
\begin{enumerate}[(A)]
 \item \emph{Selection scheme based on hole localization}\\
 In many cases, $2h1p$ and $3h2p$ ISs can be straightforwardly classified as corresponding to open or closed decay channels based on either core/valence character or spatial localization of the vacancies defining the $2h$ and $3h$ configurations, respectively. In the case of AD or DAD of a core vacancy, energetically accessible final states are typically characterized by all $2h$ or $3h$ configurations that do not contain the initial or deeper-lying vacancies. Consider, for example, decay of the Ne($1s$) vacancy. Any dicationic state containing only valence ($2s$ and $2p$) vacancies is possible final state of AD. Similarly, all valence $\Ne^{3+}$ states are accessible through DAD. Therefore, the $\Qsub$ subspace can be defined by ISs characterized by at least one $1s$ hole while all other ISs are included into the $\Psub$ subspace. Further examples are given in Supplementary material (SM).
 
 Similar strategy can be adopted when studying interatomic decay process in heteronuclear clusters where the MOs are spatially localized on specific atoms, such as
 \begin{equation}
  (A^+)^*B\rightarrow\left\{\begin{array}{ll}
                             A^+ + B^+ + e^- & \text{ICD} \\
                             A + B^{2+}+e^- & \text{ETMD}\cite{Zobeley2001jcp}
                            \end{array}\right.
 \end{equation}
 Here, the final states are distinguished by at least one vacancy being localized on the initially neutral cluster constituent B. Therefore, the $\Qsub$ subspace is spanned by $2h1p$ and $3h2p$ ISs characterized by all holes being localized on the initially ionized subunit $A$. In this way, the intra-atomic relaxation and correlation effects inside the subunit $A$ are taken into account in the discrete state, whereas any kind of interatomic decay can be described only through coupling to the complementary $\Psub$ subspace.
 
 \item \emph{Selection scheme based on adapted ISs}\\
 The above scheme relies on the fact that there is a direct one-to-one correspondence between the $2h1p$- and $3h2p$-like ISs and the open or closed channels of the decay process. In a most general situation, this is not the case. As an example, consider ICD in neon dimer,
 \begin{equation}
  \Ne_2^+(2\sigma_{g/u}^{-1})\rightarrow \Ne^+ + \Ne^+ + e^-.
 \end{equation}
The final states of the decay process are of two-site character, with each hole localized on a different atom. However, Ne$_2$ MOs are delocalized over both atoms due to inversion symmetry. In turn, $2h$ configurations derived from those MOs, such as $3\sigma_g^{-1}3\sigma_g^{-1}$, are neither two- nor one-site (both holes localized on the same atom) and the corresponding $2h1p$ ISs cannot be directly associated with either $\Psub$ or $\Qsub$ subspace. The issue can be resolved by a localization procedure described in Ref.~\onlinecite{Averbukh2006jcp} and generalized in Ref.~\onlinecite{Kolorenc2015jcp} as follows.

To restore the localized (one- or two-site) character of the $2h$ configurations, correlated $2h$ wave functions can be constructed through diagonalization of the lowest-order Hamiltonian matrix corresponding to the doubly-ionized system,
\begin{equation}
 \label{eq:2hHam}
 H_{ij,i',j'} = \langle\Phi_0|c_{j'}^\dag c_{i'}^\dag H c_i c_j|\Phi_0\rangle.
\end{equation}
The same procedure can be applied directly to ISs. First, the $2h1p$ ISs are divided into subsets characterized by the particle orbital $p$. When the corresponding small blocks of the ADC Hamiltonian $\mathbf{M}$ are diagonalized, the spectrum reflects that of the matrix \eqref{eq:2hHam} with the eigenvalues being essentially only uniformly shifted by the presence of an extra electron in the virtual orbital $p$. Corresponding eigenstates represent \emph{adapted} ISs which can be directly associated with the correlated $2h$ wave functions and, in turn, with open or closed decay channels. The adapted ISs can therefore be readily sorted into $\Psub$ and $\Qsub$, respectively. Identical procedure can be applied also to $3h2p$ space -- adapted ISs associated with correlated $3h$ wave functions are obtained by diagonalization of ADC Hamiltonian sub-blocks corresponding to $3h2p$ ISs characterized by the pair of virtual orbitals $p$, $p'$. 
\end{enumerate}

The classification scheme B is in principle completely general. The only input required for the calculations is the number of di- and tri-cationic states accessible in the decay process. For each virtual orbital $p$, corresponding number of lowest-lying adapted $2h1p$ ISs is included in the $\Psub$ subspace, and similarly for the $3h2p$ ISs. However, special care is needed if the energy gap dividing open and closed channels is very narrow. The extra electron in virtual orbital can then lead to strong mixing of open and closed channels in the adapted $2h1p$ ISs, breaking the strictly bound character of the $\Qsub$ subspace. In such a case, it might be necessary to further restrict this subspace by excluding the affected ISs. It is also advisable to associate the adapted ISs with the decay channels by comparing its $2h$ component with the eigenvectors of \eqref{eq:2hHam} rather than to rely solely on the energy ordering as the extra electron can swap closely-lying levels. The same holds also for the $3h2p$ class of adapted ISs. The principal advantage of the scheme A is, on the other hand, its simplicity. In general, it is advisable to compare both schemes whenever applicable in order to verify reliability of the results.

Once the $\Qsub$ and $\Psub$ subspaces are defined within the full configuration space, the initial state of the decay process is represented by a discrete state $|\Phi\rangle$ selected among the eigenstates of the Hamiltonian matrix $\mathbf{QMQ}$ projected onto the $\Qsub$ subspace. The selection criterion is typically the leading $1h$ configuration. The decay continuum spanned by the components $|\chi_{\beta,\epsilon}\rangle$ is approximated by eigenstates $|\chi_i\rangle$ corresponding to the discrete eigenvalues $\epsilon_i$ of the $\mathbf{PMP}$ Hamiltonian projected onto the $\Psub$ subspace.

\chng{
Computational cost of the Fano-ADC(2,2) method with selection scheme A scales formally as the standard ADC(2,2) calculations, i.e., $n_{occ}^3n_{virt}^4$ for \adcf{}. Separation into the $\Qsub$ and $\Psub$ subspaces does not involve any transformation of the ISs basis and thus merely reduces the number of occupied orbitals active in each subspace. Diagonalization of the two projected matrices is typically cheaper than diagonalization of the full matrix, but it should be noted that Fano-ADC calculations often require considerably larger basis sets than standard calculations of IPs.

Selection scheme B is more expensive as the subspaces are defined through a basis set transformation. It is not advisable to explicitly evaluate the projected $\mathbf{QMQ}$ and $\mathbf{PMP}$ matrices as they do not inherit the sparse character of $\mathbf{M}$ and the two projectors but rather to apply the three matrices sequentially in each matrix $\times$ vector operation. The whole procedure thus consists of one full $n_{occ}^3n_{virt}^4$ multiplication and two projections of $n_{occ}^3 n_{virt}^2 n_{3h}$ complexity, where $n_{3h}$ is the number of closed or open triply ionized decay channels for the $\Qsub$ and $\Psub$ subspace, respectively. $n_{3h}$ scales approximately as cube of the core or valence orbitals, but in a typical calculation the inequality $n_{virt}^2\gg n_{3h}$ holds and the relative cost of the projections is negligible. It should also be noted that even though full matrix $\times$ vector operation is performed, the projected matrices being diagonalized are smaller, resulting in a lower number of iterations required to reach the desired accuracy.
}

\chng{
\subsection{Stieltjes imaging}
}
The discrete character of the $\mathbf{PMP}$ spectrum prevents straightforward use of the eigenfunctions $|\chi_i\rangle$ as an approximation of the background continuum functions in the decay width formula \eqref{eq:GammaE}. First, these wave functions do not satisfy correct scattering boundary conditions and are normalized to unity rather than energy,
\begin{equation}
    \langle\chi_i|\chi_j\rangle = \delta_{ij}.
\end{equation}
Second, the discretized spectrum has to be interpolated in order to evaluate the resonance width $\Gamma=\Gamma(E_\Phi)$ at the desired energy as the condition $\epsilon_i=E_\Phi$ is in general not fulfilled for any of the discrete levels, except by a coincidence.

Both issues can be efficiently resolved using the so-called Stieltjes imaging \cite{Langhoff1979empmc,Hazi1979empmc} technique. The approach relies on the fact that while the wave functions $|\chi_i\rangle$ cannot be used to evaluate the decay width function \eqref{eq:GammaE} directly, they provide good approximations of its spectral moments,
\begin{equation}
    \label{eq:GammaSk}
    S_k = \int E^k \Gamma(E) dE = 2\pi\sum_{\beta=1}^{N_c}\int E^k |\langle\Phi|H-E|\chi_{\beta,E-E_\beta}\rangle|^2 dE \approx 2\pi\sum_i (\epsilon_i)^k |\langle\Phi|H-E|\chi_i\rangle|^2.
\end{equation}
Here, we have used the assumption that, within the region defined by the spatial extent of the discrete state $|\Phi\rangle$, the solutions $|\chi_i\rangle$ can replace the basis formed by the exact background continuum wave functions,
\begin{equation}
    \label{eq:discretized_unity}
    \sum_{\beta=1}^{N_c}\int |\chi_{\beta,\epsilon_\beta}\rangle\langle\chi_{\beta,\epsilon_\beta}| \approx \sum_i |\chi_i\rangle\langle\chi_i|.
\end{equation}
Using the lowest $2n_S$ spectral moments \eqref{eq:GammaSk} ($k=0,\dots,2n_S-1$), an approximation of order $n_S$ of the decay width function can be recovered using the moment theory. At each order, the decay width is obtained in terms of the $n_S$-point integration quadrature with \emph{a priori} unknown weight function $\Gamma(E)$. An efficient implementation of the Stieltjes imaging procedure is described in Ref.~\onlinecite{Muller1989pra}. In this approach, negative moments $S_{-k}$ are used to improve numerical stability. Convergence of the calculations can be controlled by performing series of approximations of increasing order $n_S$.

It follows from the Eq.\ \eqref{eq:discretized_unity} that it is not possible to formulate a rigorous procedure for the calculation of the partial decay widths $\Gamma_\beta(E)$. This problem is common to all $\lsq$ methods as the decay channels are defined only asymptotically with respect to the position of the outgoing particle and, therefore, true continuum functions are needed. However, partial decay widths can still be estimated by constructing approximate channel projectors $P_\beta$ in terms of the $\lsq$ ISs and repeating the Stieltjes imaging procedure with projected functions $P_\beta|\chi_i\rangle$. The method is detailed in Refs.~\onlinecite{Cacelli1986mp,Averbukh2005jcp}. In the present work, we only use this approach to estimate the three-electron collective Auger decay branching ratio in Kr in Sec.\ \ref{sec:KrCAD}. In this particular case, the $\Psub$ subspace is spanned by $2h1p$ ISs only and the definition of projectors corresponding to two- and three-electron decay pathways is straightforward. General discussion of partial widths in the framework of Fano-ADC(2,2) method will be subject of a future publication.

\section{Results and discussion}
\label{sec:Results}

\subsection{Ionization energies}
\label{sec:IPs}

Before applying the Fano-ADC(2,2) method to calculations of intra- and inter-atomic electronic decay widths, we will demonstrate the accuracy of the ADC(2,2) approximation scheme for ionization energies. Since the main goal of the work is to develop a method which treats the main $1h$ and satellite $2h1p$ states consistently, we are interested particularly in the relative energy positions of the two classes of states. In this section, we present results of benchmark calculations of atomic and molecular vertical ionization potentials (IP), which can be directly compared to available experimental and theoretical data. In Tabs.\ \ref{tab:BeIP} and \ref{tab:NeIP}, energies of main and satellite ionization states of Be and Ne atoms, computed using different variants of ADC(2,2) scheme, are compared with standard CI-SD and ADC(2)x methods. Values from the NIST database \cite{NIST_ASD} are used as reference.
\begin{table}[ht]
 \begin{tabular}{l|c|c|c|ccc}
  configuration & NIST & CI-SD & ADC(2)x & \adcf & \adcx & \adcm \\
  \hline
  \multicolumn{7}{c}{frozen $1s$ core} \\
  \hline
  $1s^2 2s^2$ GS & & -397.81 & \multicolumn{4}{c}{-397.37 (MBPT2)} \\
  \hline
  $1s^2 2s$ & \textbf{9.32} & 9.30$_{-0.02}$ & 8.83$_{-0.49}$ & 8.92$_{-0.40}$ & 8.86$_{-0.46}$ & 8.86$_{-0.46}$ \\
  $1s^2 2p$ & \textbf{13.28} & 13.29$_{+0.01}$ & 12.03$_{-1.25}$  & 12.85$_{-0.43}$ & 12.85$_{-0.43}$ & 12.85$_{-0.43}$ \\
  $1s^2 3d$ & \textbf{21.48} & 21.38$_{-0.10}$ & 20.12$_{-1.36}$ &  20.94$_{-0.54}$ & 20.94$_{-0.54}$ & 20.94$_{-0.54}$ \\
  $1s^2 4s$ & \textbf{23.64} & 23.54$_{-0.10}$ & 22.28$_{-1.35}$ &  23.10$_{-0.54}$ & 23.10$_{-0.54}$ & 23.10$_{-0.54}$ \\
  \hline
  \multicolumn{7}{c}{$1s$, $2s$ active} \\
  \hline
  $1s^2 2s^2$ GS & & -398.51 & \multicolumn{4}{c}{-398.08 (MBPT2)} \\
  \hline
  $1s^2 2s$ & \textbf{9.32} & 9.26$_{-0.06}$ & 8.87$_{-0.46}$ & 8.95$_{-0.37}$ & 8.90$_{-0.43}$ & 8.90$_{-0.43}$ \\
  $1s^2 2p$ & \textbf{13.28} & 13.84$_{+0.55}$ & 12.01$_{-1.27}$ & 12.82$_{-0.46}$ & 12.82$_{-0.46}$ & 12.83$_{-0.45}$ \\
  $1s^2 3d$ & \textbf{21.48} & 22.03$_{+0.55}$ & 20.12$_{-1.36}$ & 21.01$_{-0.47}$ & 21.01$_{-0.47}$ & 21.01$_{-0.47}$ \\
  $1s^2 4s$ & \textbf{23.64} & 24.19$_{+0.55}$ & 22.28$_{-1.36}$ & 23.16$_{-0.48}$ & 23.17$_{-0.47}$ & 21.17$_{-0.47}$ \\
  \hline
 \end{tabular}
 \caption{\label{tab:BeIP}Be ionization potentials (in eV). Lower indices at the computed energies indicate differences from the reference NIST values. In the upper part of the table, results obtained with frozen $1s$ core orbital in all post-HF methods are given, while for the lower part, both $1s$ and $2s$ orbitals were kept active. Neutral ground state energies were computed using corresponding CI-SD and MBPT2 methods with the same sets of active orbitals.}
\end{table}

For Tab.\ \ref{tab:BeIP}, IPs of Be were computed using the aug-cc-pV5Z basis set further augmented by $4s4p4d$ Rydberg-like Gaussian functions\cite{Kaufmann1989jpb}. Upper half shows results obtained with the $1s$ orbital frozen in the post-HF methods (i.e., only $2s$ vacancies were allowed in the configuration spaces). The data show significant improvement of the $2h1p$ satellite states IPs obtained using ADC(2,2) in comparison to the ADC(2)x method. IP of the $1s^22s^1$ main state remains basically unchanged. As a result, the whole spectrum is consistently shifted by about 0.5\,eV towards lower energies relatively to the experimental benchmark, regardless the character of the state. In this particular example, all variants of ADC(2,2) yield equivalent results.

For only one active occupied orbital, CI-SD method comprises complete expansion and is, therefore, clearly superior over ADC. It should be noted, however, that to obtain IPs shown in Tab.\ \ref{tab:BeIP}, independent CI-SD calculation of the neutral ground state energy has to be performed. In contrast, in ADC schemes the ground state correlation is included directly in the secular matrix $\mathbf{M}$ through the higher-order matrix elements. Therefore, ionization energies are obtained through a single matrix diagonalization. MBPT2 ground state energies are given for completeness and do not enter the calculation of IPs.

Second half of Tab.\ \ref{tab:BeIP} demonstrates size-extensivity of the ADC method. In Be, energy separation of $1s$ and $2s$ orbitals is about 120\,eV and, therefore, core and valence excitations constitute essentially separated subsystems. Consequently, ionization energies computed using the ADC methods remain unaltered when the $1s$ orbital is included into the active space. CI-SD, on the other hand, is no longer full expansion and the inconsistency of the resulting truncated scheme is reflected in lower accuracy of the satellite states energies.

\begin{table}[ht]
 \begin{tabular}{l|c|c|c|ccc}
  configuration & NIST & CI-SD & ADC(2)x & \adcf & \adcx & \adcm \\
  \hline
  $2s^2 2p^6$ GS & & -3506.09 & \multicolumn{4}{c}{-3506.33 (MBPT2)} \\
  \hline
  $2s^2 2p^5\,{}^2P$ & \textbf{21.56} & 21.69$_{+0.13}$ & 20.73$_{-0.83}$ & 21.48$_{-0.09}$ & 20.23$_{-1.34}$ & 20.14$_{-1.43}$ \\
  $2s^1 2p^6\,{}^2S$ & \textbf{48.48} & 48.78$_{+0.30}$ & 47.46$_{-1.01}$ & 48.19$_{-0.28}$ & 46.59$_{-1.88}$ & 46.43$_{-2.05}$ \\
  $2s^2 2p^4 3s\,{}^2P$ & \textbf{49.35} & 52.64$_{+3.29}$ & 56.40$_{+7.05}$ & 48.43$_{-0.92}$ & 48.43$_{-0.92}$ & 45.75$_{-3.60}$ \\
  $2s^2 2p^4 3s\,{}^2D$ & \textbf{52.11} & 55.45$_{+3.34}$ & 58.26$_{+6.15}$ & 51.10$_{-1.01}$ & 51.10$_{-1.01}$ & 48.29$_{-3.82}$ \\
  $2s^2 2p^4 3p\,{}^2D$ & \textbf{52.69} & 55.89$_{+3.21}$ & 59.07$_{+6.38}$ & 51.77$_{-0.91}$ & \chng{51.77$_{-0.91}$} & 48.67$_{-4.02}$ \\
  $2s^2 2p^4 3p\,{}^2S$ & \textbf{52.91} & 56.08$_{+3.18}$ & 59.30$_{+6.39}$ & 51.96$_{-0.94}$ & 51.96$_{-0.94}$ & 48.78$_{-4.12}$ \\
  \hline
 \end{tabular}
 \caption{\label{tab:NeIP}Ne ionization potentials (in eV) computed using the aug-cc-pV5Z basis set further augmented by $4s4p4d$ Rydberg-like Gaussian functions\cite{Kaufmann1989jpb} with frozen $1s$ core orbital. Lower indices at the computed energies indicate differences from the reference NIST values.}
\end{table}
Tab.\ \ref{tab:NeIP} shows lowest six IPs of Ne atom computed by the same methods, employing aug-cc-pV5Z basis set further augmented by $4s4p4d$ Rydberg-like Gaussian functions\cite{Kaufmann1989jpb}. In this case, both ADC(2)x and CI-SD are clearly inadequate to describe correlation in satellite states, yielding IPs by more than 6\,eV and 3\,eV too high, respectively. The minimal \adcm{} scheme provides some improvement of the satellite states over ADC(2)x but the magnitude of the error is still large, comparable to CI-SD. Accuracy of the main state IPs is even lower than at the ADC(2)x level. Moreover, in contrast to CI-SD and ADC(2)x, the $2h1p$ states IPs are significantly underestimated to the extent that the lowest $2s^22p^43s\,{}^2P$ satellite state falls below the $2s^1 2p^6\,{}^2S$ main state. Going to the \adcx{} scheme leads to considerably higher accuracy of the satellite state IPs. Main state IPs, on the other hand, are virtually unchanged compared to \adcm{} and show even larger errors than the satellite states.

\chng{Tab.\ \ref{tab:NeIP} might suggest that \adcx{} provides the most balanced treatment of main and satellite states. However, application to molecules shows that the main states energies in both \adcx{} and \adcm{} schemes are not shifted uniformly but rather determined with random errors as large as 1.5\,eV. This inconsistency is introduced by the first order $2h1p/3h2p$ coupling and is only corrected by the second order $1h/2h1p$ matrix elements in the \adcf{} scheme, which yields $1h$ main state IPs comparable to CI-SD and clearly superior to ADC(2)x.}

In Tab.\ \ref{tab:22fvs3}, we compare main states vertical IPs of selected small closed-shell molecules computed using the hierarchy of ADC(2), ADC(2)x, \adcf{} and ADC(3)\cite{Trofimov2005jcp} approximation schemes. The \adcf{} calculations were performed using aug-cc-pVDZ basis set (cc-pVDZ for H) as in Ref.~\onlinecite{Trofimov2005jcp}, all other data are taken from Tab.~V therein (including the experimental values). The comparison shows that \adcf{} main state energies are of similar quality as in the third-order method. Compared to the ADC(3) scheme, only the third-order $M^{(3)}_{11}$ contribution to the $1h/1h$ block is missing in \adcf{}, which at least for the listed molecules does not play significant role. Of course, when only main ionization state energies are required, ADC(3) scheme is clearly superior due to much smaller explicit configuration space, spanned only by the $1h$ and $2h1p$ ISs.
\begin{table}[ht]
 \begin{tabular}{lr|c|c|c|c|c}
  \multicolumn{2}{c|}{vacancy} & Expt. & ADC(2) & ADC(2)x & ADC(3) & \adcf \\
  \hline
  C$_2$H$_4$ & $1b_{2u}$ & \textbf{10.95} & 10.15$_{-0.80}$ & 10.09$_{-0.86}$ & 10.46$_{-0.49}$ & 10.45$_{-0.50}$ \\
  & $1b_{2g}$ & \textbf{12.95} & 12.79$_{-0.16}$ & 12.57$_{-0.38}$ & 13.19$_{+0.25}$ & 12.91$_{-0.04}$ \\
  & $3a_{g}$ & \textbf{14.88} & 13.79$_{-1.09}$ & 13.67$_{-1.21}$ & 14.36$_{-0.52}$ & 14.71$_{-0.17}$ \\
  & $1b_{3u}$ & \textbf{16.34} & 16.13$_{-0.21}$ & 15.61$_{-0.73}$ & 16.49$_{+0.15}$ & 15.85$_{-0.49}$ \\
   & & \textbf{17.80} & & 18.08$_{+0.28}$ & 18.12$_{+0.32}$ & \chng{17.43$_{-0.37}$} \\
  & $2b_{1u}$ & \textbf{19.40}  & 18.96$_{-0.44}$ & 18.08$_{-1.32}$ & 19.00$_{-0.40}$ & 18.84$_{-0.56}$ \\
  & & \textbf{20.45} & & 19.92$_{-0.53}$ & 20.02$_{-0.43}$ & 19.49$_{-0.96}$ \\
  CO & $5\sigma$ & \textbf{14.01}  & 13.78$_{-0.23}$ & 13.43$_{-0.58}$ & 13.80$_{-0.21}$ & 14.02$_{+0.01}$ \\
   & $1\pi$ & \textbf{16.91}  & 16.23$_{-0.68}$ & 16.30$_{-0.61}$ & 16.88$_{-0.03}$ & 16.82$_{-0.09}$ \\
   & $4\sigma$ & \textbf{19.72} & 18.30$_{-1.42}$ & 18.42$_{-1.30}$ & 20.10$_{+0.38}$ & 19.42$_{-0.30}$ \\
  F$_2$ & $1\pi_g$ & \textbf{15.80} & 13.88$_{-1.92}$ & 13.97$_{-1.83}$ & 15.87$_{+0.07}$ & 15.47$_{-0.33}$ \\
   & $1\pi_u$ & \textbf{18.80} & 17.03$_{-1.77}$ & 16.84$_{-1.96}$ & 19.11$_{+0.31}$ & 18.59$_{-0.21}$ \\
   & $3\sigma_g$ & \textbf{21.10} & 20.24$_{-0.86}$ & 20.48$_{-0.62}$ & 21.01$_{-0.09}$ & 20.83$_{-0.27}$ \\
  HF & $1\pi$ & \textbf{16.05} & 14.39$_{-1.66}$ & 14.93$_{-1.12}$ & 16.41$_{+0.36}$ & 15.73$_{-0.32}$ \\
   & $3\sigma$ & \textbf{20.00} & 18.67$_{-1.33}$ & 19.11$_{-0.89}$ & 20.30$_{+0.30}$ & 19.72$_{-0.28}$ \\
  N$_2$ & $3\sigma_g$ & \textbf{15.60} & 14.79$_{-0.81}$ & 14.72$_{-0.88}$ & 15.60$_{+0.00}$ & 15.78$_{+0.18}$ \\ 
   & $1\pi_u$ & \textbf{16.98} & 16.99$_{+0.01}$ & 16.90$_{-0.08}$ & 16.77$_{-0.21}$ & 17.17$_{+0.19}$ \\
   & $2\sigma_u$ & \textbf{18.78} & 17.99$_{+0.79}$ & 17.62$_{-1.16}$ & 18.93$_{+0.15}$ & 18.76$_{-0.02}$ \\
   \hline 
   \multicolumn{2}{c|}{$\bar{\Delta}_\text{abs}$}  & & 0.89 & 0.91 & 0.26 & 0.29 \\
   \multicolumn{2}{c|}{$\Delta_\text{max}$} & & 1.92 & 1.96 & 0.52 & 0.96 \\
 \end{tabular}
 \caption{\label{tab:22fvs3}Molecular vertical ionization potentials (in eV) computed using IP-EOM-CCSDT, ADC(2), ADC(3) and \adcf\ methods with aug-cc-pVDZ basis sets. ADC(2), ADC(3) and experimental values are taken from Ref.~\onlinecite{Trofimov2005jcp}. Lower indices give errors of the computed energies as compared to the experiment. In the last two lines, mean absolute error $\bar{\Delta}_\text{abs}$  and the maximum absolute error $\Delta_\text{max}$ are given.
 }
\end{table}

\chng{
Comparison between the hierarchy of ADC schemes and equation-of-motion coupled-cluster methods (IP-EOM-CC) for main ionization states can be found in Ref.~\onlinecite{Trofimov2005jcp}. For CO, F$_2$ and N$_2$, both CCSD and CCSDT are more accurate than ADC(3), but both are also more computationally expensive. CCSDT provides particularly accurate results with an average error below 0.1\,eV, but the required solution of the neutral ground state CCSDT equations scales as\cite{Musial2003jcp} $n_{occ}^3n_{virt}^5$. Comparison of satellite state IPs, computed using \adcf{}, ADC(2)x and different variants of IP-EOM-CC is shown in Tab.\ \ref{tab:CCsatellites}. CCSDTQ level is taken as reference. The computationally cheapest CCSD, scaling as $n_{occ}^2 n_{virt}^4$, is clearly insufficient for representation of satellite states. Considering solely the average error, ADC(2)x is surprisingly accurate, but it is likely coincidental (note also the large spread of errors). In comparison to the CCSDTQ reference, CCSDT and \adcf{} produce identical average absolute errors, the two methods consistently over- and underestimating the IPs, respectively. Owing to the large spread of available data, comparison to experiment is also inconclusive.
}
%
%
\begin{table}[ht]
 \chng{
 \begin{tabular}{ll|c|c|c|c|c|c}
  \multicolumn{2}{c|}{} & Exp. & CCSDTQ & CCSD & CCSDT & ADC(2)x & \adcf{} \\
  \hline
 CO & $\tilde{D}\, {}^2\Pi$ & 22.7, 22.0 &  \textbf{22.87} & 26.34$_{+3.47}$ & 23.23$_{+0.36}$ & 22.87$_{+0.00}$ & 22.19$_{-0.68}$ \\
 & $3\, {}^2\Sigma^+$ & 23.4, 23.6, 24.1 & \textbf{23.74} & 26.36$_{+2.62}$ & 24.00$_{+0.26}$ & 22.72$_{-1.02}$ & 23.28$_{-0.46}$\\
 N$_2$ & $\tilde{D}\, {}^2\Pi_g$ & 24.79, 25.0  & \textbf{24.22} & 28.27$_{+4.05}$ & 24.78$_{+0.56}$ & 24.06$_{-0.16}$  & 23.80$_{-0.42}$\\
 & $\tilde{C}\, {}^2\Sigma_u^+$ & 25.51 & \textbf{24.99} & 28.78$_{+3.79}$ & 25.28$_{+0.29}$ & 24.76$_{-0.23}$  & 24.55$_{-0.44}$\\
 & $2s\sigma\,\, {}^2\Sigma_g^+$ & & \textbf{37.96}$^a$ & 38.58$_{+0.62}^a$ & 38.58$_{+0.62}^a$ & 36.63$_{-1.33}$  & 37.65$_{-0.31}$\\
 \hline
 \multicolumn{2}{c|}{$\bar{\Delta}_\mathrm{abs}$} & & & 2.91 & 0.42 & 0.55 & 0.46 \\
 \multicolumn{2}{c|}{$\Delta_\mathrm{max}$} & & & 4.05 & 0.62 & 1.33 & 0.68
 \end{tabular}\\
 \caption{\label{tab:CCsatellites}Comparison of satellite states IPs in CO and N$_2$ computed using \adcf{} and IP-EOM-CC methods. Unless indicated otherwise by a superscript, all calculations were carried out employing cc-pVDZ basis set. CC and experimental data reprinted from Ref.~\onlinecite{Kamiya2006jcp} except the $2s\sigma$ vacancy state of N$_2$ from Ref.~\onlinecite{Matthews2016jcp}.\\
 ${}^a$ANO0 basis set\cite{Almlof1987jcp} 
  }
  }
\end{table}

To summarize, ADC(2,2) comprises significant step towards the desired balance of the treatment of main and satellite states without loosing the benefit of size-extensivity. From the point of view of the IPs only, the results of the minimal \adcm{} variant are not satisfactory. \adcx{} and \adcf{} provide the expected accuracy of the satellite state IPs -- the errors correspond in magnitude exactly to those of the main state energies in ADC(2)x. 
\chng{However, the accuracy of the main state energies deteriorate in \adcm{} and \adcx{} and must be corrected by the second-order $1h/2h1p$ couplings included in the \adcf{} scheme.}
In fact, the \adcf{} main state IPs are then comparable to that of the higher-order ADC(3) scheme. Neither of the ADC(2,2) variants thus provides perfectly balanced scheme, \chng{but at least in atoms} the relative shift of the main and satellite states IPs is reduced by nearly 90\% in \adcf{}.

\subsection{Auger decay widths}
\label{sec:Auger}
In this and the following section, we present results of application of the Fano-ADC(2,2) method to intra-atomic Auger decay widths. We focus on processes in which higher-order three-electron transitions\cite{Kolorenc2016jpb} play a measurable role. In particular, we consider double Auger decay\cite{Journel2008pra,Zeng2013pra} (DAD) of a single core vacancy, in which two electrons are ejected to continuum. As a prototype of the DAD process can be considered the production of Ne$^{3+}$ in the decay of the $1s$ core vacancy in neon\cite{Carlson1965prl},
\begin{equation}
    \label{eq:Ne1sDAD}
    \Ne^+(1s^{-1})\,\rightarrow\,\left\{\begin{array}{ll}
        \Ne^{2+} + e^- & \quad\text{AD} \\
        \Ne^{3+} +e_1^- + e_2^- & \quad\text{DAD} 
    \end{array}\right. .
\end{equation}
In this case, DAD proceeds solely simultaneously -- both secondary electrons are ejected at the same time without involvement of any intermediate state. Most recent experiments estimate the DAD branching ratio as 6\% of the total decay rate\cite{Kolorenc2016jpb}. For heavier atoms, the importance of DAD increases, together with that of the cascade decay pathway. In the cascade decay, the secondary electrons are emitted sequentially with an intermediate dicationic metastable state being populated between the two steps. For the Kr $3d$ vacancy, for instance, the DAD branching ratio was determined between 20\%\cite{Palaudoux2010pra} and 30\%\cite{Tamenori2004jpb} and the process is strongly dominated by the sequential decay.
Another relevant three-electron process is shake-up during Auger decay, represented here by the
\begin{equation}
 \label{eq:Mg2s}
 \text{Mg}^+(2s^1\,2p^6\,3s^2)\,\rightarrow\,\text{Mg}^{2+}(2s^2\,2p^5\, (3p,4s)^1)+e^-
\end{equation}
transition in the decay of the Mg $2s$ vacancy.

Tab.\ \ref{tab:DAD} lists total decay widths of Mg$^+(2s^{-1})$, Ne$^+(1s^{-1})$, Ar$^+(2p^{-1})$ and Kr$^+(3d^{-1})$ Auger-active states, obtained using different variants of the Fano-ADC method. Available experimental or theoretical values are given for comparison, together with branching ratios of the DAD or shake-up processes. For each ADC(2,2) variant, relative difference from the Fano-ADC(2)x result is evaluated,
\begin{equation}
    \label{eq:GammaDelta}
    \Delta = (\Gamma_\text{ADC(2,2)}-\Gamma_\text{ADC(2)x})/\Gamma_\text{ADC(2,2)}.
\end{equation}
The uncertainties given for Fano-ADC values are determined as statistical standard deviations related solely to the Stieltjes imaging procedure -- decay widths are evaluated through averaging of nine consecutive orders $n_S$ of the imaging procedure in the region of best convergence. Hence, the error margins do not attempt to reflect any systematical errors connected with the Fano-ADC methodology.
\begin{table}[ht]
{\footnotesize
 \begin{tabular}{l|cc|c|cc|cc|cc}
   \multicolumn{1}{c|}{}& \multicolumn{2}{c|}{Reference} & \multicolumn{1}{c|}{ADC(2)x} & \multicolumn{2}{c|}{\adcf} & \multicolumn{2}{c|}{\adcx} & \multicolumn{2}{c}{\adcm} \\
   & $\Gamma$\,(meV) & DAD/shake-up & $\Gamma$\,(meV) & $\Gamma$\,(meV) & $\Delta$ & $\Gamma$\,(meV) & $\Delta$ & $\Gamma$\,(meV) & $\Delta$ \\
   \hline
   \hline
   Mg$^+$($2s^{-1}$) & 680\cite{Kochur2001aa} & 13\% & $585\pm3$ & $621\pm9$ & $(6\pm 2)\%$ & $684\pm 7$ & $(14\pm 1)\%$ & $674\pm 7$ & $(13\pm 1)\%$ \\
   \hline
   Ne$^+$($1s^{-1}$) & $257\pm6$\cite{Mueller2017aj} & 6\%\cite{Kolorenc2016jpb} & $244\pm 4$ & $271\pm 5$ & $(10\pm 3)\%$ & $318\pm 4$ & $(23\pm 2)\%$ & $283\pm 10$ & $(14\pm 4)\%$ \\
   \hline
   Ar$^+$($2p^{-1}$) & $125\pm 5$\cite{Zeng2013jpb} & $(13\pm 2)\%$\cite{Zeng2013jpb} & $114 \pm 5$ & $125\pm 3$ & $(10\pm 6)\%$ & $149\pm 4$ & $(24\pm 5)\%$ & $134\pm 4$ & $(18\pm 6)\%$ \\
   \hline
   Kr$^+$($3d^{-1}$) & $88\pm 4$\cite{Jurvansuu2001pra} & 20-30\%\cite{Tamenori2004jpb,Palaudoux2010pra} & $68\pm 2$ & $94\pm 2$ & $(27\pm 4)\%$ & $109\pm 2$ & $(37\pm 4)\%$ & $96\pm 3$ & $(29\pm 4)\%$ \\
   \hline
 \end{tabular}
}
 \caption{\label{tab:DAD}Total Auger decay widths and DAD/shake-up contributions available in literature (second and third column) for atomic core vacancies. Fourth column shows total decay widths computed by the Fano-ADC(2)x method, columns 5-10 total contain decay widths obtained using variants of Fano-ADC(2,2) and their relative differences $\Delta$ from Fano-ADC(2)x values. For details on the basis sets and the $\mathcal{Q/P}$ classification procedure, see text and SM.
}
\end{table}

Compared to the reference values, Fano-ADC(2)x method underestimates the decay widths in all listed cases. Using the new \adcf{} scheme, the agreement is improved significantly, in particular for the Ar and Kr atoms. For Ne, \adcf{} somewhat overestimates the reference value. However, considering the error margins, the agreement is in fact good for both ADC(2)x and \adcf{} schemes. Only for the Mg$(2s^{-1})$ vacancy, even the \adcf{} value stays underestimated by nearly $9\%$ with respect to the reference.
Other two variants, \adcm{} and \adcx{}, yield total decay widths systematically larger than \adcf{} (on average by 9\% and 20\%, respectively). With the exception of  Mg$^+(2s^{-1})$ vacancy this leads to substantially overestimated results compared to the reference values.

The errors of the ADC(2)x results with respect to the reference values correspond approximately to the DAD or shake-up contributions. The differences \eqref{eq:GammaDelta} therefore provide good estimates of the three-electron branching ratios and can be used as indication to what extent the higher-order processes are accounted for in the ADC(2,2) schemes. The results show that for DAD, particularly the \adcf{} scheme performs very well. It should be pointed out, however, that the difference $\Delta$ is a composite effect of
\chng{the DAD contribution and improved description of the single AD transitions due to better representation of both the initial $1h$-like and final $2h1p$-like states.}
Partial decay widths determined using suitable channel projectors\cite{Averbukh2005jcp} are needed for reliable analysis of the DAD and shake-up contributions and will be discussed in a follow-up work.

The agreement between our results and the literature decay width of Mg$^+(2s^{-1})$ having a significant contribution of the shake-up process is less satisfactory. In this case, \adcx{} provides apparently the best result, but the agreement is likely incidental. Shake-up transitions are, in principle, more difficult to describe properly within Fano-ADC method than the DAD process. Considering the particular transition of Eq.\ \eqref{eq:Mg2s}, $3h2p$ ISs deriving from the $(2p^{-1}\,3s^{-2})$ $3h$ configurations belong to both $\mathcal{P}$ and $\mathcal{Q}$ subspaces to account for the open $\text{Mg}^{2+}(2s^2\,2p^5\, \{3p,4s\}^1)$ shake-up final channels and closed $\text{Mg}^{3+}(2s^2\,2p^5)$ triply-ionized channels, respectively. In the framework of a $\mathcal{L}^2$ method, however, it is not possible to rigorously distinguish between those two types of final states. In the present calculations, $3h2p$ ISs derived from $2p^{-1}\,3s^{-2}$ configurations were included only into the $\mathcal{P}$-subspace as it is paramount in the Fano theory to fully eliminate continuum from the $\mathcal{Q}$ subspace. It may come at the expense of missing some correlation in the initial state, which might be the reason for the worse agreement with literature values. It should be also pointed out that another source of uncertainty lies in the literature value. It was computed \emph{ab initio} using the $R$-matrix method\cite{Kochur2001aa} and, to the best of our knowledge, was not verified experimentally. The present calculations call for revisiting the Mg$^+(2s^{-1})$ decay width both theoretically and experimentally. 

To conclude, Fano-\adcf{} method represents substantial improvement over the original Fano-ADC(2)x approach. \adcm{} and in particular \adcx{} variants are inferior, confirming the inconsistencies observed already in the ionization energies of Ne and the arguments motivating the \adcf{} scheme given in Sec.\ \ref{sec:ADC22}.

\subsection{Collective three-electron Auger decay}
\label{sec:KrCAD}

Another relevant three-electron decay process is the so-called collective Auger decay (CAD) of two vacancies, which recombine simultaneously to eject single secondary electron. To this date, the strongest known atomic three-electron CAD is the relaxation of the $\text{Kr}^+(3d^{10}\,4s^{0}\,4p^6\,5p)$ excited state of ionized Kr, which was conclusively observed in a cascade process initiated by $3d\rightarrow 5p$ excitation of Kr,
\begin{equation}
 \text{Kr}^*(3d^9\,4s^2\,4p^6\,5p)\,\rightarrow\,\text{Kr}^+(3d^{10}\,4s^{0}\,4p^6\,5p)+e^-\,\rightarrow\,\left\{
 \begin{array}{ll}
  \text{Kr}^{2+}(3d^{10}\,4s^1\,4p^5)+e^- & \text{AD} \\
  \text{Kr}^{2+}(3d^{10}\,4s^2\,4p^4)+e^- & \text{CAD}
 \end{array}
 \right. .
\end{equation}
In the coincidence experiment carried out by Eland \emph{et al}\cite{Eland2015njp}, the three-electron process was found to be about 40 times weaker than the competing two-electron process, corresponding to branching ratio of 2.4\%. As such, decay of the $\text{Kr}^+(3d^{10}\,4s^{0}\,4p^6\,5p)$ state is perfect case to test the new Fano-ADC(2,2) method.

In terms of excitation classes, the above CAD process corresponds to a $2h1p\,\rightarrow\,2h1p$ transition and can be described already in the Fano-ADC(2)x method, at least in principle. The calculations were carried out using the same basis set as for the Kr$^+(3d^{-1})$ decay in previous section (see SM). To correctly determine the CAD contribution, partial decay widths are needed. In this particular case, however, the problem can be handled already at the present stage of development. Since no $3h2p$ ISs enter the final $\mathcal{P}$ subspace, the approach previously implemented\cite{Averbukh2005jcp,Kolorenc2015jcp,Stumpf2017chp} within the Fano-ADC(2)x framework can be directly applied. In particular, projector onto the AD channel is defined by $2h1p$ ISs containing the $4s$ hole while the CAD channel is defined by two $4p$ vacancies.

Results are collected in Tab.\ \ref{tab:KrCAD}. While Fano-ADC(2)x estimates the three-electron branching ratio to be only 0.2\%, which account for only about 8\% of the measured intensity, Fano-\adcf{} scheme recovers 88\% of the experimental value. While \adcx{} scheme leads to essentially the same results as the \adcf{}, \adcm{} in this case fails. Specifically, the spectrum of the projected $QHQ$ Hamiltonian does not show the expected structure of $\text{Kr}^+(3d^{10}\,4s^{0}\,4p^6\,nl)$ states, therefore, it is not possible to identify unambiguously the resonance of interest among the $QHQ$ eigenvectors.

The total decay widths calculated using the ADC(2)x and \adcf{} schemes differ by a factor of 2.2. Such a large difference cannot be attributed to the higher-order transitions but rather to the poor representation of the $2h1p$-like decaying state in the ADC(2)x scheme. Direct comparison with experiment is not available. Owing to the low resolution of the experimental data, it is not possible to determine reliably the lifetime broadening of the observed spectral lines. A rough estimate suggests the upper limit of about 250\,meV, indicating that even the \adcf{} result might be still too high.
\begin{table}[ht]
  \begin{tabular}{lcc}
   method & $\quad\Gamma$\,(meV)$\quad$ & CAD BR \\
   \hline
   experiment\cite{Eland2015njp} & - & 2.4\% \\
   ADC(2)x & 744 & 0.2\% \\
   \adcf{} & 331 & 2.1\%
  \end{tabular}
 \caption{\label{tab:KrCAD} Calculated decay widths of the $\text{Kr}^+(3d^{10}\,4s^{0}\,4p^6\,5p)$ state and three-electron CAD branching ratios.}
\end{table}

\subsection{Interatomic decay widths}
\label{sec:ICD}
In this section, we demonstrate the applicability of Fano-ADC(2,2) method to calculation of interatomic decay rates. We will focus on simple, previously studied problems, namely ICD of the $2s$ vacancy in $\Ne_2$ and of the $(1s^{-2}\,2p^1)$ resonances in ionized-excited $\He_2$ dimers. Characteristic feature of these dipole-allowed ICD transitions is the $R^{-6}$ dependence\cite{Santra2001prb,Gokhberg2010pra} of the decay widths on the interatomic distance $R$, valid for large separations. The asymptotic decay widths can then be well approximated by the virtual photon transfer model due to Matthew and Komninos\cite{Matthew1975surfsci,Averbukh2004prl},
\begin{equation}
 \label{eq:vphm}
 \Gamma(R) = \frac{3\hbar}{4\pi}\left(\frac{c}{\omega}\right)^4\frac{\tau_\text{rad}^{-1}\,\sigma}{R^6},
\end{equation}
where $\tau_\text{rad}$ is the radiative lifetime of the initial vacancy in isolated donor atom and $\sigma$ is the ionization cross section (at the virtual photon energy $\hbar\omega$) of the acceptor atom. Reproducing this behavior is therefore fundamental test of the method.

\begin{figure}[ht]
 \includegraphics[width=\linewidth]{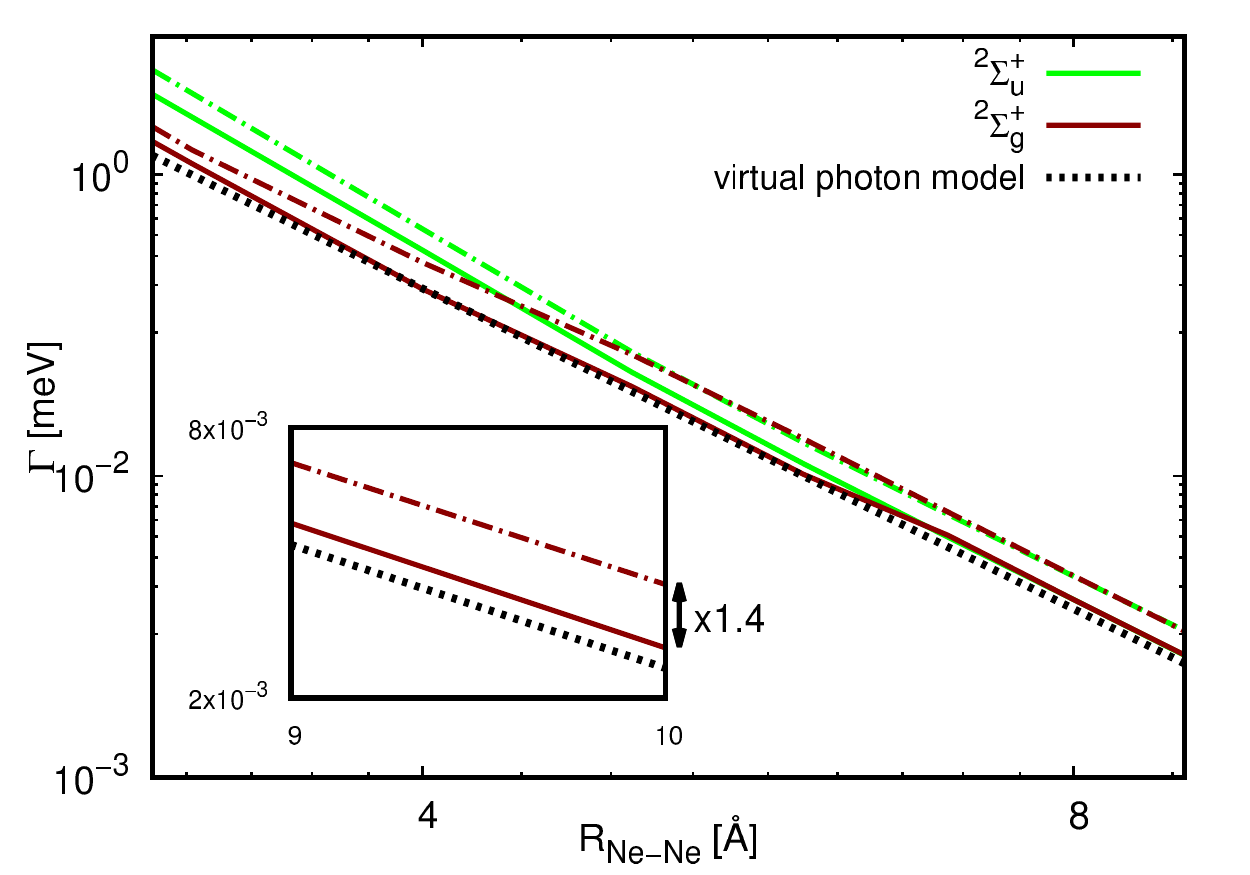}
 \caption{\label{fig:Ne2ICD}Doubly-logarithmic plot of total ICD widths of the $(2\sigma_g^{-1})\,{}^2\Sigma_g^+$ (dark red) and $(2\sigma_u^{-1})\,{}^2\Sigma_u^+$ (light green) vacancy states in Ne$_2$, calculated using Fano-\adcf{} (full lines) and Fano-ADC(2)x (dashed-dotted lines). Dotted line corresponds to the virtual photon model, Eq.\ \eqref{eq:vphm} (atomic data for Ne are taken from Refs.~\onlinecite{Griffin2001jpb,Samson2002jesrp}). Inset shows magnified comparison the \emph{ab initio} methods and virtual photon model at large interatomic separation.}
\end{figure}

Decay of Ne $2s$ vacancy in neon dimer,
\begin{equation}
 \label{eq:Ne2ICD}
 \Ne^+(2s^{-1})\Ne\,\rightarrow\,\Ne^+(2p^{-1})+\Ne^+(2p^{-1})+e^-,
\end{equation}
is one of the most thoroughly investigated examples of ICD, both theoretically\cite{Santra2000prl,Scheit2004jcp, Averbukh2006jcp} and experimentally\cite{Jahnke2004prl,Schnorr2013prl}. Due to the inversion symmetry of the homonuclear dimer, the $2s$ vacancy split into a pair of states delocalized over the whole dimer, $(2\sigma_g^{-1})\,{}^2\Sigma_g^+$ and $(2\sigma_u^{-1})\,{}^2\Sigma_u^+$. Fig.\ \ref{fig:Ne2ICD} shows dependence of the total ICD widths on the internuclear distance $R$ for both gerade and ungerade states, calculated using Fano-\adcf{} (full lines) and Fano-ADC(2)x (dahed-dotted lines). Details about the present calculations (basis sets, $\QP$ partitioning) can be found in SM.

Qualitatively, both Fano-\adcf{} and Fano-ADC(2)x yield very similar results. The decay width of the gerade initial state follows the $R^{-6}$ trend over the whole range of internuclear distances while that of the ungerade state is somewhat enhanced at short distances (by a factor of about 2 at the equilibrium interatomic distance). Quantitatively, however, Fano-\adcf{} decay widths are approximately by a factor of 1.4 smaller for both initial states and over the whole range of interatomic distances. Comparison with Eq.\ \eqref{eq:vphm} at large interatomic distances (see inset of Fig.\ \ref{fig:Ne2ICD}) indicate that Fano-\adcf{} results are more accurate. Asymptotically, decay widths obtained by Fano-ADC(2)x results are by a factor of 1.5 larger than the prediction of virtual photon model. This discrepancy was attributed to the inaccuracies of the ADC(2)x theoretical description\cite{Averbukh2006jcp}. Indeed, improved description of the $2h1p$-like final states reduces this error by more than 70\%.

\begin{figure}[ht]
\includegraphics[width=\linewidth]{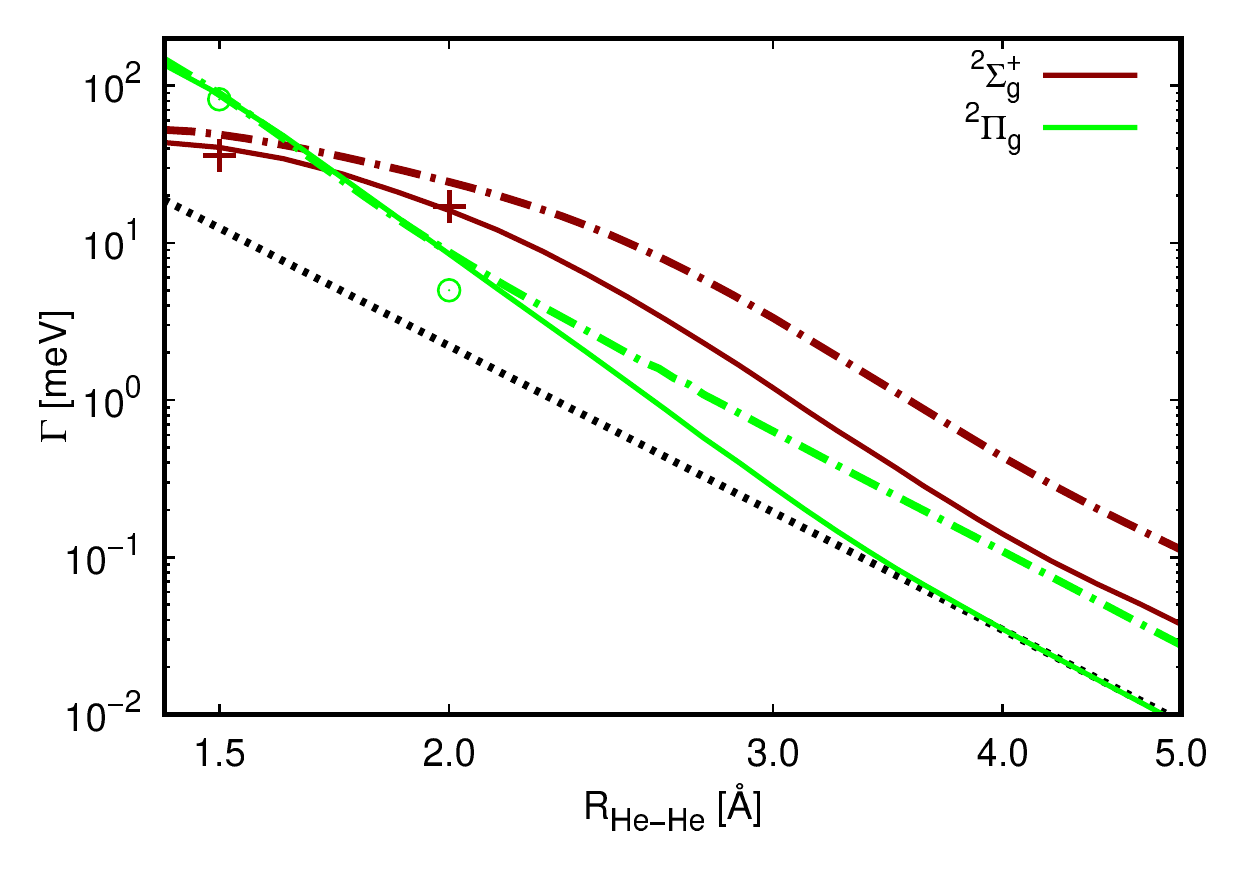}
\caption{\label{fig:He2g}Total ICD widths of the ${}^2\Sigma_g$ (dark red) and ${}^2\Pi_g$ (light green) states of the $\He^+(1s^0\,2p^1)$ type. Results fo Fano-\adcf{} and Fano-ADC(2)x methods are shown by full and dashed-dotted lines, respectively. Results of $R$-matrix calculations\cite{Sisourat2017jcp} are indicated by crosses (${}^2\Sigma_g$) and circles (${}^2\Pi_g$). $R^{-6}$ dependence (dotted line) is fitted to the ${}^2\Pi_g$ state to guide the eye.}
\end{figure}
\begin{figure}[ht]
\includegraphics[width=\linewidth]{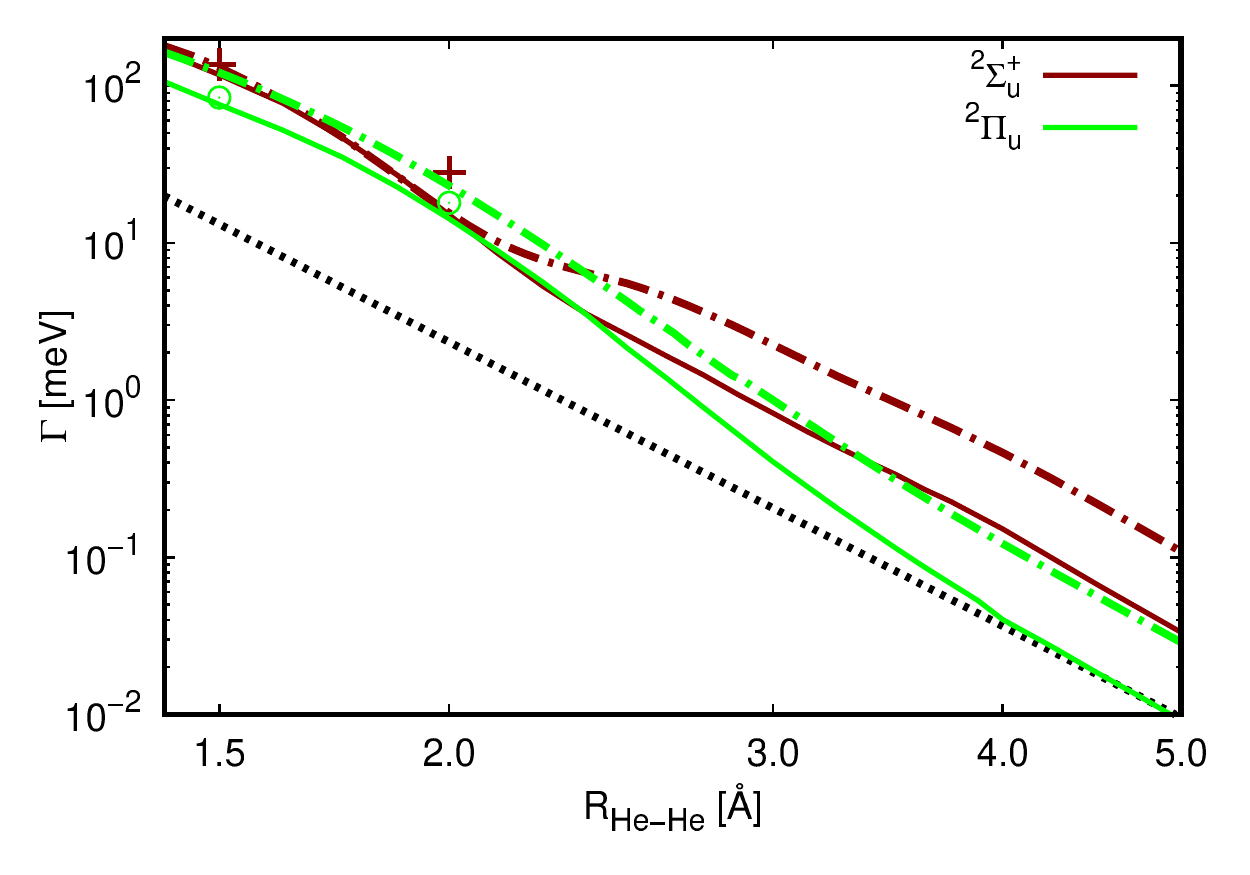}
\caption{\label{fig:He2u}Total ICD widths of the ${}^2\Sigma_u$ (dark red) and ${}^2\Pi_u$ (light green) states of the $\He^+(1s^0\,2p^1)$ type. Results fo Fano-\adcf{} and Fano-ADC(2)x methods are shown by full and dashed-dotted lines, respectively. Results of $R$-matrix calculations\cite{Sisourat2017jcp} are indicated by crosses (${}^2\Sigma_u$) and circles (${}^2\Pi_u$). $R^{-6}$ dependence (dotted line) is fitted to the ${}^2\Pi_u$ state to guide the eye.}
\end{figure}
\begin{figure}
    \centering
    \includegraphics[width=\linewidth]{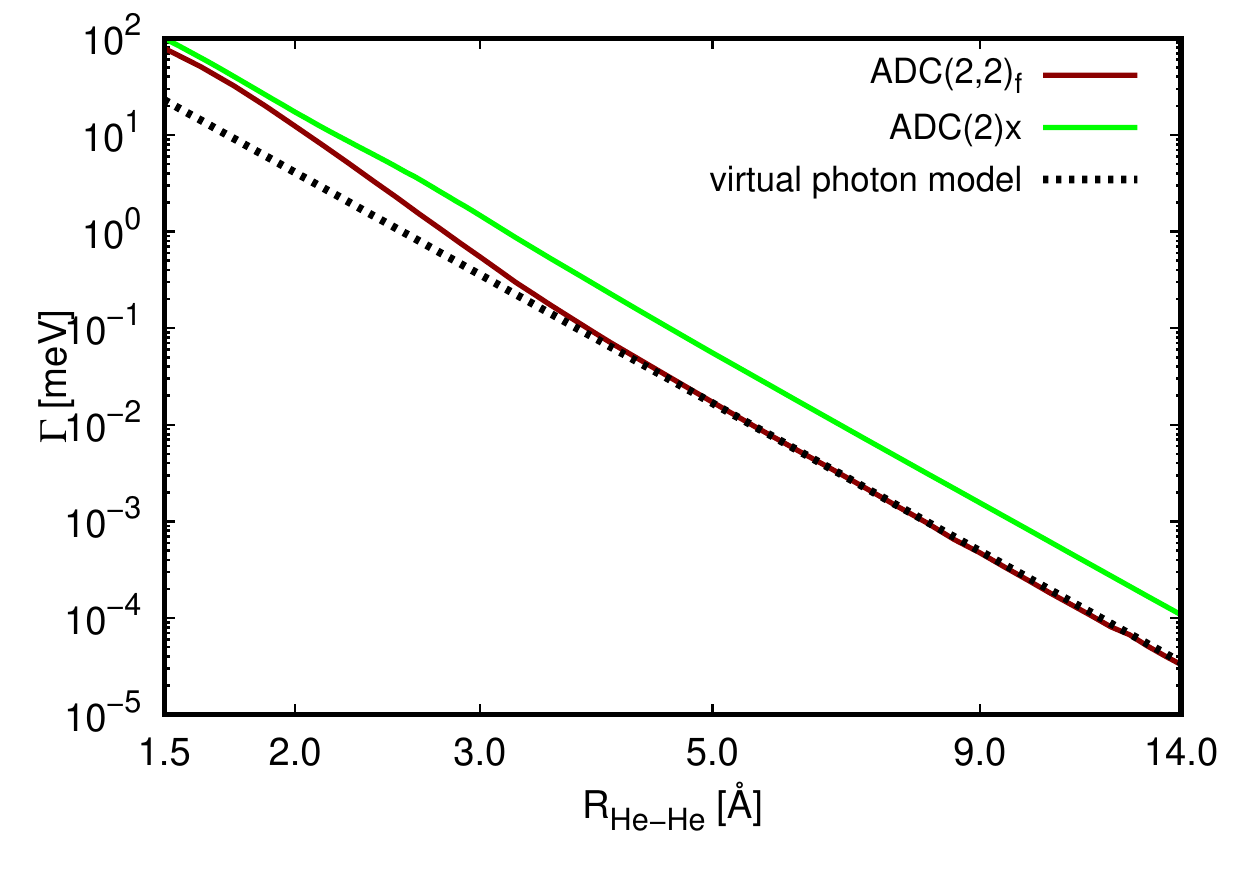}
    \caption{\label{fig:He2virt}Comparison of the state-averaged ICD decay widths \eqref{eq:GammaHe2_aver} for $\He_2$ with the virtual photon model \eqref{eq:vphm}. Radiative lifetime $\tau$ of the $\He^+(2p^1)$ state and the He photoionization cross section $\sigma$ are taken from Refs.~\onlinecite{Marr1976adndt,Drake1992pra}.}
\end{figure}
ICD in helium dimer,
\begin{equation}
\label{eq:He2ICD}
 \He^+(1s^0\,2p^1)\He\,\rightarrow\,\He^+(1s)+\He^+(1s)+e^-,
\end{equation}
is another thoroughly researched interatomic decay process. Its experimental realization\cite{Havermeier2010prl,Sisourat2010nphys} even provided direct visualization of the nodal structure of the vibrational wave function of the decaying state. Corresponding decay widths were studied extensively using Fano-ADC(2)x method\cite{Kolorenc2010pra,Kolorenc2015jcp} and more recently also with $R$-matrix\cite{Sisourat2017jcp}. As in the case of neon dimer, the initial metastable states are delocalized over the dimer, giving rise to gerade-ungerade pairs derived from atomic $2p$ orbitals oriented parallel (${}^2\Sigma_g$, ${}^2\Sigma_u$) and perpendicular (${}^2\Pi_g$, ${}^2\Pi_u$) to the dimer axis. Decay widths of the gerade states are shown in Fig. \ref{fig:He2g}, of the ungerade states in Fig.\ \ref{fig:He2u}. Results of Fano-\adcf{} and Fano-ADC(2)x are shown as full and dashed-dotted lines, respectively. Values obtained using the $R$-matrix method at the CI-SD level (cf.\ scheme 2 in Ref.~\onlinecite{Sisourat2017jcp}) are shown at the two interatomic distances at which they were computed. Details of the present calculations are given in SM.

For both methods, the decay widths follow the expected $R^{-6}$ trend for internuclear separations larger than 4\,\AA. The $\Gamma_\Sigma/\Gamma_\Pi$ ratio also quickly approaches the value of 4, which can be deduced from the dipole orientation of the states involved in the transition\cite{Gokhberg2010pra}. Quantitatively, however, the two methods differ considerably more than in the case of neon dimer. Comparison with the virtual photon model \eqref{eq:vphm}, which is in this case provided by the state-averaged decay width,
\begin{equation}
    \label{eq:GammaHe2_aver}
    \bar{\Gamma}=\frac{1}{6}(\Gamma_{\Sigma_g}+2\Gamma_{\Pi_g}+\Gamma_{\Sigma_u}+2\Gamma_{\Pi_u}),
\end{equation}
is shown in Fig.\ \ref{fig:He2virt}. While the \adcf{} result agrees for $R>5\,\text{\AA}$ with the asymptotic formula within 5\%, ADC(2)x yields decay widths by a factor of 3.1 too large. Towards smaller interatomic separations, the difference decreases as the Fano-\adcf{} decay widths are significantly enhanced relative to the $R^{-6}$ trend while the Fano-ADC(2)x deviate much less. As a result, the level of agreement with the $R$-matrix calculations\cite{Sisourat2017jcp} below $R=2\,\text{\AA}$ is similar for both methods.
The large asymptotic discrepancy is to be attributed to the $2h1p\,\rightarrow\,2h1p$ character of the interatomic transition. First order representation provided by the  ADC(2)x scheme for both the initial and final states is clearly insufficient in this case.

\section{Conclusions}
\label{sec:conclusions}

In the framework of the well-established Fano-ADC methodology, we have presented a method of computation of intra- and interatomic nonradiative decay widths based on the Fano theory of resonances and a new ADC(2,2) scheme for representation of the many-electron wave functions. The principal asset of the new scheme is the balanced representation of both initial and final states of the decay processes, where both are being treated correctly up to the second order of perturbation theory.

The ability of the new method to provide accurate decay widths is demonstrated on a number of Auger decay and ICD processes. In the case of ICD, superiority of the Fano-ADC(2,2) method is exemplified by near-perfect agreement of the \emph{ab initio} decay widths with the virtual photon transfer model in the region of large interatomic distances. Fano-ADC(2,2) method allows us, for the first time within the Fano-ADC approach, to take into account the higher-order three-electron transitions, such as DAD, which contribute significantly to the total decay widths and represent basic manifestations of electron correlation\cite{Feifel2018SciRep}. Comparison of our results to the available benchmarks indicates that Fano-ADC(2,2) provides quantitatively correct description of such higher-order transitions.

\chng{
The increased accuracy and broader capability of Fano-ADC(2,2) comes at a price, namely the significant growth of computational demands connected with the $3h2p$ excitation class. Compared to Fano-ADC(2)x, the computational cost increases from $n_{occ}^3 n_{virt}^2$ to $n_{occ}^3 n_{virt}^4$. The use of the new method is thus likely to be limited to small systems when high accuracy is required or to investigate higher-order decay processes while Fano-ADC(2)x remains the method of choice for larger polyatomic systems.
}

\section*{Supplementary material}
See supplementary material for information about basis sets and other computational details.

\begin{acknowledgments}
Financial support by the Czech Science Foundation (Project GA\v{C}R No.\ 17-10866S) is gratefully acknowledged. V.A. acknowledges support by EPSRC/DSTL MURI grant EP/N018680/1.
\end{acknowledgments}

\section*{Data availability}
The data that support the findings of this study are available from the corresponding author upon reasonable request.

\appendix

\section{Spin-orbital form of ISR-ADC(2,2) working equations}
\label{app:ADC22equations}
In this appendix, we list explicit expressions (in spin-orbital form) for elements of the ISR-ADC(2,2) secular matrix
\begin{equation*}
 \mathbf{M} = \mathbf{M}^{(0)}+\mathbf{M}^{(1)}+\mathbf{M}^{(2)}.
\end{equation*}
With the exception of the second-order contribution to the $2h1p/2h1p$ block, they are equal to the previously published non-Dyson ADC(3) scheme \cite{Schirmer1998jcp} [note the sign convention $\mathbf{M}=-(\mathbf{K}+\mathbf{C})$].
%
%
We use the short notation
\begin{equation}
 v_{abij}=\frac{V_{ab[ij]}}{\epsilon_a+\epsilon_b-\epsilon_i-\epsilon_j}
\end{equation}
with $V_{ab[ij]}=V_{abij}-V_{abji}=\langle ab||ij\rangle$ being the antisymmetrized Coulomb integral in the ``1212'' convention and $\epsilon_p$ denote the HF orbital energy. The letters $i,j,k,l,\dots$ and $a,b,c,\dots$ reffer to the occupied and unoccupied orbitals, respectively.
\begin{enumerate}
 \item $1h/1h$ block
 \begin{equation}
  \label{eq:Mkk0}
  M_{kk'}^{(0)} = -\epsilon_k \delta_{kk'}
 \end{equation}
 \begin{equation}
  M_{kk'}^{(1)} = 0
 \end{equation}
 \begin{equation}
  M_{kk'}^{(2)} = \frac{1}{2}\sum_{a,b,j} v_{abkj} v_{abk'j}^*\left(\epsilon_a+\epsilon_b-\epsilon_j-\frac{1}{2}(\epsilon_k+\epsilon_{k'})\right)
 \end{equation}
 \item $1h/2h1p$ block
 \begin{equation}
  M_{i,akl}^{(1)} = V_{kl[ia]}
 \end{equation}
 \begin{equation}
  M_{i,akl}^{(2)} = \frac{1}{2}\sum_{b,c} v_{bckl}^* V_{bc[ai]}-\left[\sum_{b,j}v_{ablj}^*V_{kb[ij]}\right]+[k\leftrightarrow l]
 \end{equation}
 \item $2h1p/2h1p$ block
 \begin{equation}
  M_{akl,a'k'l'}^{(0)} = (\epsilon_a-\epsilon_k-\epsilon_l)\delta_{aa'}\delta_{kk'}\delta{ll'}
 \end{equation}
 \begin{equation}
  M_{akl,a'k'l'}^{(1)} = \delta_{aa'} V_{k'l'[kl]}-\left[\delta_{kk'} V_{l'a[la']}+\delta_{ll'} V_{k'a[ka']}\right]+[k\leftrightarrow l]
 \end{equation}
 \begin{equation}
  M_{akl,a'k'l'}^{(2)} = M_{akl,a'k'l'}^{(A)}+M_{akl,a'k'l'}^{(B)}+M_{akl,a'k'l'}^{(C)}+M_{akl,a'k'l'}^{(D)}+M_{akl,a'k'l'}^{(E)},
 \end{equation}
 where
 \begin{multline}
  M_{akl,a'k'l'}^{(A)} = \left[\frac{1}{2}\delta_{aa'}\delta_{kk'}\sum_{b,c,j}v_{bclj}v_{bcl'j}^*\left(\epsilon_b+\epsilon_c-\epsilon_j-\frac{1}{2}(\epsilon_l+\epsilon_l')\right)\right]\\
  +[k\leftrightarrow l,k'\leftrightarrow l']-[k\leftrightarrow l]-[k'\leftrightarrow l'],
 \end{multline}
 \begin{equation}
  M_{akl,a'k'l'}^{(B)} = \frac{1}{2}\delta_{kk'}\delta_{ll'}\sum_{c,i,j}v_{acij}v_{a'cij}^*\left(\epsilon_c-\epsilon_i-\epsilon_j+\frac{1}{2}(\epsilon_a+\epsilon_{a'})\right),
 \end{equation}
 \begin{equation}
  M_{akl,a'k'l'}^{(C)} = -\frac{1}{2}\delta_{aa'}\sum_{b,c}v_{bckl}v_{bck'l'}^*\left(\epsilon_b+\epsilon_c-\frac{1}{2}(\epsilon_k+\epsilon_{k'}+\epsilon_l+\epsilon_{l'})\right),
 \end{equation}
 \begin{multline}
  M_{akl,a'k'l'}^{(D)} = \left[-\delta_{kk'}\sum_{c,j}v_{aclj}v_{a'cl'j}\left(\epsilon_c-\epsilon_j+\frac{1}{2}(\epsilon_a+\epsilon_{a'}-\epsilon_l-\epsilon_{l'})\right)\right] \\
  +[k\leftrightarrow l,k'\leftrightarrow l']-[k\leftrightarrow l]-[k'\leftrightarrow l'],
 \end{multline}
 \begin{equation}
  \label{eq:Maklakl2E}
  M_{akl,a'k'l'}^{(E)}=\sum_c v_{ackl}v_{a'ck'l'}^*\left(\epsilon_c+\frac{1}{2}(\epsilon_a+\epsilon_{a'}-\epsilon_k-\epsilon_{k'}-\epsilon_l-\epsilon_{l'})\right).
 \end{equation}
 Note that in $M_{akl,a'k'l'}^{(B)}$, the possible contribution proportional to $\delta_{kl'}\delta_{lk'}$, corresponding to $(k'\leftrightarrow l')$ permutation, is missing due to the $k<l$, $k'<l'$ restriction on the spin-orbitals defining permissible IS's (cf.\ Eq.\ \eqref{eq:CJ}).
 \chng{These restrictions also eliminate number of possible permutations in the expressions below.}
 \chng{
 \item $2h1p/3h2p$ block
 \begin{equation}
     M_{akl,a'b'k'l'm'}^{(1)} = M_{akl,a'b'k'l'm'}^{(A)} + M_{akl,a'b'k'l'm'}^{(B)},
 \end{equation}
 where
 \begin{equation}
     M_{akl,a'b'k'l'm'}^{(A)} = [\delta_{kk'}\delta_{ll'}V_{m'a[b'a']}] - [m'\leftrightarrow k'] + [k'l'm'\leftrightarrow l'm'k'],
 \end{equation}
 \begin{multline}
     M_{akl,a'b'k'l'm'}^{(B)} = \left\{\delta_{aa'}\left([\delta_{kk'}V_{l'm'[b'l]}] - [k\leftrightarrow l] - [k'\leftrightarrow l'] + [k\leftrightarrow l,k'\leftrightarrow l']\right.\right. \\ 
     \left.\left.+ [k'l'm'\leftrightarrow m'k'l']-[k\leftrightarrow l,k'l'm'\leftrightarrow m'k'l]\right)\right\} - \{a'\leftrightarrow b'\}.
 \end{multline}
 \item $3h2p/3h2p$ block
 \begin{equation}
     M_{abklm,a'b'k'l'm'}^{(0)} = (\epsilon_a+\epsilon_b-\epsilon_k-\epsilon_l-\epsilon_m)\delta_{aa'}\delta_{bb'}\delta_{kk'}\delta_{ll'}\delta_{mm'}
 \end{equation}
 \begin{equation}
     M_{abklm,a'b'k'l'm'}^{(1)} = M_{abklm,a'b'k'l'm'}^{(A)} + M_{abklm,a'b'k'l'm'}^{(B)} + M_{abklm,a'b'k'l'm'}^{(C)},
 \end{equation}
 where
 \begin{equation}
     \label{eq:3h2p3h2p_worst}
     M_{abklm,a'b'k'l'm'}^{(A)} = \delta_{kk'}\delta_{ll'}\delta_{mm'}V_{ab[a'b']},
 \end{equation}
 \begin{multline}
     M_{abklm,a'b'k'l'm'}^{(B)} = \left\{\delta_{aa'}\left([\delta_{kk'}\delta_{ll'}V_{m'b[b'm]}] - [l\leftrightarrow m] - [l'\leftrightarrow m']+[l\leftrightarrow m,l'\leftrightarrow m']\right.\right. \\
     +[klm\leftrightarrow lmk]-[klm\leftrightarrow lmk,l'\leftrightarrow m']+[k'l'm'\leftrightarrow l'm'k'] \\
     \left.\left.- [m\leftrightarrow l,k'l'm'\leftrightarrow l'm'k'] +[klm\leftrightarrow lmk,k'l'm'\leftrightarrow l'm'k'] \right) \right\} \\
     -\{a\leftrightarrow b\}-\{a'\leftrightarrow b'\}+\{a\leftrightarrow b,a'\leftrightarrow b'\},
 \end{multline}
 \begin{multline}
     M_{abklm,a'b'k'l'm'}^{(C)} = \delta_{aa'}\delta_{bb'}\left( [\delta_{kk'}V_{l'm'[lm]}] - [k\leftrightarrow l]+[klm\leftrightarrow mkl] \right. \\
     -[k'\leftrightarrow l']+[k\leftrightarrow l,k'\leftrightarrow l'] - [klm\leftrightarrow mkl,k'\leftrightarrow l'] + [k'l'm'\leftrightarrow m'k'l'] \\
     \left. -[k\leftrightarrow l,k'l'm'-m'k'l'] + [klm\leftrightarrow mkl,k'l'm'\leftrightarrow m'k'l']\right)
 \end{multline}
}
\end{enumerate}

The explicit expressions for the secular matrix as given above are formulated with respect to electron configurations in the spin-orbital form, defined in Eq.\ \eqref{eq:CJ} ("primitive" excitations). For an efficient implementation it is necessary to generate spin-free working equations for the total spin value of interest (typically $S=1/2$ in order to couple with the $1h$ states). This is achieved by standard angular momentum algebra techniques. First, spin-adapted excitations are formed as appropriate linear combinations of the primitive excitations. Application of the corresponding unitary transformation to the secular matrix defined by Eqs.\ \eqref{eq:Mkk0}-\eqref{eq:Maklakl2E} then leads to decoupled block structure of the matrix corresponding to different values of $S$. Subsequently, spin summations in the acquired PT expressions for the matrix elements can be performed, yielding the spin-free formulas in terms of only spatial two-electron integrals and HF orbital energies.


\bibliography{papers}

\end{document}


\title{Fano-ADC(2,2) method for electronic decay rates -- supplementary material} 

\author{P.\ Koloren\v{c}}
\affiliation{Charles University, Faculty of Mathematics and Physics, Institute of Theoretical Physics, V Hole\v{s}ovi\v{c}k\'ach 2, 180 00 Prague, Czech Republic}
\author{V.\ Averbukh}
\affiliation{Blackett Laboratory, Imperial College London, London SW9 2AZ, UK}

\maketitle 

\section*{Basis sets and other computational details}
%
Here we provide information on the basis sets and other details of calculations reported in the present work. All basis sets were downloaded from the basis set exchange library \cite{pritchard2019a}. MOLCAS\cite{molcas74} package was used for HF calculation and Coulomb integral evaluations.

\subsection*{Sec.\ IV\,B, Tab.\ V}
\begin{itemize}
    \item Mg$^+(2s^{-1})$\\[0.4em]
    \underline{basis:} uncontracted cc-pCV5Z\cite{feller1996a,schuchardt2007a} augmented by  10s10p8d continuum-like  Gaussian basis functions of the Kaufmann-Baumeister-Jungen (KBJ) type\cite{Kaufmann1989jpb}\\[0.4em]
    \underline{configuration space restrictions:} $1s$ orbital frozen, virtual orbitals restricted to energies below 80\,Ha\\[0.4em]
    \underline{$Q/P$ selection scheme A:} $\Psub$ subspace spanned by $2h1p$ ISs with at least one $3s$ hole and by $3h2p$ ISs with two $3s$ holes, with the exception of ISs with at least one $2s$ vacancy, which belong naturally to the $\Qsub$ subspace
    \item Ne$^+(1s^{-1})$\\[0.4em]
    \underline{basis:} uncontracted aug-cc-pCV5Z\cite{dunning1989jcp,kendall1992jcp} augmented by 10s10p8d even-tempered Gaussian functions\\[0.4em]
    \underline{configuration space restrictions:} virtual orbitals restricted to energies below 460\,Ha\\[0.4em]
    \underline{$Q/P$ selection scheme A:} all ISs characterized by at least one $1s$ vacancy belong to the $\Qsub$ subspace
    \item Ar$^+(2p^{-1})$ \\[0.4em]
    \underline{basis:} uncontracted aug-cc-pwCV5Z\cite{peterson2002jcp,woon1993jcp} augmented by 4s3p even-tempered Gaussian functions\\[0.4em]
    \underline{configuration space restrictions:} $1s$ orbital frozen,  virtual orbitals restricted to energies below 30\,Ha\\[0.4em]
    \underline{$Q/P$ selection scheme A:} ISs characterized by at least one $2s$ or $2p$ vacancy belong to the $\Qsub$ subspace
    \item Kr$^+(3d^{-1})$ \\[0.4em]
    \underline{basis:} cc-pwCV5Z-PP\cite{peterson2003jcp,peterson2010jcp} (with  Stuttgart-Koeln MCDHF RSC ECP for [Ne]), augmented by 5s5p5d2f continuum-like KBJ Gaussian functions; H- and I-type functions removed from the basis\\[0.4em]
    \underline{configuration space restrictions:} $3s$ and $3p$ orbitals frozen, virtual orbitals restricted to energies below 11\,Ha \\[0.4em]
    \underline{$Q/P$ selection scheme A:} $\Qsub$ subspace is defined by configurations with at least one $3d$ vacancy plus all $3h2p$ ISs associated with Kr$^+(4s^{-2}\,4p^{-1}\,nl^{+2})$ configurations
\end{itemize}

Convergence of the decay widths with respect to basis (in particular, the augmentation and virtual orbitals restriction) was determined using the Fano-ADC(2)x scheme -- extensive convergence studies based on the ADC(2,2) schemes are impractical due to the large $3h2p$ class of intermediate states. The approach was tested by running Fano-\adcf{} calculations for Kr($3d^{-1}$) and Mg($2s^{-1}$) using two different augmentations and without any restriction on virtual orbitals. Choice between even-tempered and KBJ-type augmentation is based on the quality of convergence of the Stieltjes imaging procedure.

In Tab.\ V, $Q/P$ selection scheme A was used for all atoms. Tab.\ \ref{tab:DAD_QPAB} compares the Fano-\adcf{} results with decay widths obtained using scheme B with the same number of open decay channels. Both schemes yield results which agree within the error margins deduced from the convergence of the Stieltjes imaging procedure.
\begin{table}[ht]
    \begin{tabular}{c|c|c}
     & scheme A & scheme B \\[0.4em]
   \hline
   \hline
   Mg$^+$($2s^{-1}$) & $621\pm9$ & $618\pm 3$ \\[0.4em]
   \hline
   Ne$^+$($1s^{-1}$) & $271\pm 5$ & $272\pm 4$ \\[0.4em]
   \hline
   Ar$^+$($2p^{-1}$) & $125\pm 3$ & $127\pm 1$ \\[0.4em]
   \hline
   Kr$^+$($3d^{-1}$) & $94\pm 2$ & $93\pm 2$ \\[0.4em]
   \hline
    \end{tabular}
    \caption{\label{tab:DAD_QPAB} Comparison of total Auger decay widths computed using the Fano-\adcf{} method employing $Q/P$ selection schemes A and B.}
\end{table}

\subsection*{Sec.\ IV\,C, Tab.\ VI}
Decay widths of the $\text{Kr}^+(3d^{10}\,4s^{0}\,4p^6\,5p)$ resonance were computed using $\mathcal{Q/P}$ classification scheme $B$ with the $\mathcal{P}$-subspace constructed from the adapted $2h1p$ ISs associated with the $\text{Kr}^{2+}(3d^{10}\,4s^1\,4p^5\,{}^1P,{}^3P)$ and $\text{Kr}^{2+}(3d^{10}\,4s^2\,4p^4\,{}^1S,{}^1D,{}^3P)$ open channels. All $3h2p$ ISs are included in the $\mathcal{Q}$ subspace. Same basis set and configuration restrictions as in Sec.\ IV\,B were employed.

 \subsection*{Sec.\ IV\,D}
 \begin{itemize}
  \item ICD in neon dimer (Fig.\ 1)\\[0.4em]
  \underline{basis:} cc-pVTZ augmented by $(5s,5p,5d)$ continuum-like KBJ Gaussian functions on atomic centers, with additional sets of $(4s,4p,4d)$ continuum-like KBJ Gaussian functions placed at three ghost centers - midpoint between the atoms and $\pm 3R/2$, where $R$ is the internuclear distance\\[0.4em]
  \underline{configurations space restriction:} $1s$ orbital frozen\\[0.4em]
  \underline{$Q/P$ selection scheme B:} $\Psub$ subspace comprises all adapted $2h1p$ ISs associated with the 9 singlet and 9 triplet two-site states derived (in terms of AOs) from the $\Ne(2p^{-1})-\Ne(2p^{-1})$ configurations
  \item ICD in helium dimer (Figs.\ 2-4)\\[0.4em]
  \underline{basis:} cc-pV6Z basis augmented by $(9s,9p,9d)$ continuum-like KBJ Gaussian functions on atomic centers, with additional set of $(5s,5p,5d)$ continuum-like KBJ functions at the midpoint between the atoms\\[0.4em]
  \underline{configurations space restriction:} virtual orbitals restricted to energies below 20\,Ha
  \underline{$Q/P$ selection scheme B:} $\Psub$ subspace comprises adapted $2h1p$ ISs associated with the two (${}^1\Sigma_g^+$ and ${}^3\Sigma_u^+$) $\He^+(1s)\He^+(1s)$ open channels. All $3h2p$ ISs were included into both subspaces in order to ensure equivalent correlation description of initial and final states of the process, leading to non-orthogonal $Q$ and $P$ projectors.
 \end{itemize}

\bibliography{papers}